\documentclass[%
 reprint,
superscriptaddress,
 amsmath,amssymb,
 aps,
]{revtex4-2}

\usepackage{graphicx}
\usepackage{dcolumn}
\usepackage{bm,}
\usepackage{dcolumn}
\usepackage{array,multirow}
\usepackage{hyperref}
\usepackage{xcolor}
\usepackage{braket}
\usepackage{subfiles}

\begin{document}

\preprint{APS/123-QED}

\title{Impact of atomic reconstruction on optical spectra of twisted TMD homobilayers}

\author{Joakim Hagel}
  \email{joakim.hagel@chalmers.se}
  \affiliation{%
Department of Physics, Chalmers University of Technology, 412 96 Gothenburg, Sweden\\
}%
\author{Samuel Brem}%
\affiliation{%
 Department of Physics, Philipps University of Marburg, 35037 Marburg, Germany\\
}%
\author{Johannes Abelardo Pineiro}
  \affiliation{%
Department of Physics, Philipps University of Marburg, 35037 Marburg, Germany\\
}%
  \author{Ermin Malic}%
  \affiliation{%
 Department of Physics, Philipps University of Marburg, 35037 Marburg, Germany\\
}%
\affiliation{%
Department of Physics, Chalmers University of Technology, 412 96 Gothenburg, Sweden\\
}%

\begin{abstract}
Twisted bilayers of transition metal dichalcogenides (TMDs) have revealed a rich exciton landscape including hybrid excitons and spatially trapped moir\'e excitons that dominate the optical response of the material. Recent studies have shown that in the low-twist-angle regime, the lattice undergoes a significant relaxation in order to minimize local stacking energies. Here, large domains of low energy stacking configurations emerge, deforming the crystal lattices via strain and consequently impacting the electronic band structure. However, so far the direct impact of atomic reconstruction on the exciton energy landscape and the optical properties has not been well understood. Here, we apply a microscopic and material-specific approach and predict a significant change in the potential depth for moir\'e excitons in a reconstructed lattice, with the most drastic change occurring in naturally stacked TMD homobilayers. We show the appearance of multiple flat bands and a significant change in the position of trapping sites compared to the rigid lattice. Most importantly, we predict a multi-peak structure emerging in optical absorption of WSe$_2$ homobilayers - in contrast to the single peak that dominates the rigid lattice. This finding can be exploited as an unambiguous signature of atomic reconstruction in optical spectra of moir\'e excitons in naturally stacked twisted homobilayers.
\end{abstract}
\maketitle

\section{Introduction}
The recent advances in the field of two-dimensional materials have opened up a new platform to study many-particle physics in nanomaterials \cite{ares2022recent}. The field of twistronics is of particular interest, where the relative twist angle between two vertically stacked 2D materials can be used as an additional external knob to tune the material properties \cite{ciarrocchi2022excitonic,andrei2020graphene, tong2017topological}. Consequently, the introduction of a twist angle in these layered systems can lead to intriguing many-particle phenomena, such as unconventional superconductivity and correlated phases \cite{cao2018unconventional,wang2020correlated}. Transition metal dichalcogenides (TMDs) have emerged as a significant and noteworthy subclass of two-dimensional materials, garnering substantial interest in recent years \cite{geim2013van,liao2019van}. Here, strongly bound electron-hole pairs, known as excitons, form due to the two-dimensional confinement and the reduced screening of the Coulomb interaction. They dominate the material's optical response even at room temperature \cite{perea2022exciton,mschmitt2022formation,mueller2018exciton,splendiani2010emerging}. Furthermore, long-lived interlayer excitons can form, which exhibit an out-of-plane dipole moment due to the spatial separation of electrons and holes \cite{wang2018electrical,altaiary2022electrically,huang2022spatially,rivera2015observation,miller2017long,kunstmann2018momentum,PhysRevResearch.3.043217,tagarelli2023electrical}. Additionally, the large wave function overlap can lead to a significant carrier tunneling between the layers, giving rise to hybrid states of intra- and interlayer excitons \cite{PhysRevResearch.3.043217,D0NR02160A,gerber2019interlayer,alexeev2019resonantly,mschmitt2022formation,meneghini2022ultrafast}. 

When introducing a twist angle between two vertically stacked TMDs, a periodic superlattice emerges. This leads to a moir\'e pattern, i.e. a periodic superlattice potential that can trap excitons at specific high-symmetry sites \cite{seyler2019signatures,tran2019evidence,tong2020interferences,yu2017moire,merkl2020twist,brem2020tunable,hagel2022electrical,D0NR02160A}. Furthermore, the emerging moir\'e potential can be tuned by changing the twist angle which has been shown to have a large impact on the optical response of these materials \cite{seyler2019signatures,tran2019evidence,choi2021twist,merkl2020twist,Forg2021,jin2019observation}. Interestingly, recent studies have revealed that when the twist angle is small, i.e. the period of the moir\'e pattern is large, the two monolayer crystal lattices do no longer remain rigid, but instead undergo a deformation to reduce the total adhesion energy \cite{weston2020atomic,rosenberger2020twist,van2023rotational,naik2022intralayer,li2021imaging,zhang2017interlayer,Zhao2023,Li2023,sung2020broken}. Here, some energetically favorable high-symmetry sites grow to form large domains, separated by typically narrow domain walls (cf. \autoref{fig:1}). It has been predicted that this atomic reconstruction should have a major impact on the electronic band structure and fundamentally change the potential landscape \cite{andersen2021excitons,enaldiev2020stacking,naik2022intralayer,carr2018relaxation,ferreira2021band}. However, it is still not thoroughly understood how atomic reconstruction can be identified in the optical response of moir\'e excitons. 

In this work, we investigate photoluminescence (PL) and absorption spectra in atomically reconstructed TMD bilayers. Our approach is based on the ab initio derived continuum model for the deformation of the rigid lattice derived in Refs. \cite{enaldiev2020stacking,carr2018relaxation} combined with an equation-of-motion approach for the optical response of moir\'e excitons \cite{PhysRevResearch.3.043217,hagel2022electrical,brem2020tunable,D0NR02160A}. This allows us to obtain microscopic insights into the changes in the moir\'e exciton energy landscape and the optical response in the presence of atomic reconstruction. The deformation of the moir\'e supercell into large periodic domains induces strain that can result in deeper potentials for moir\'e excitons. As a direct consequence, we find more trapped states than in the rigid lattice and the stacking region in which excitons are trapped can differ from the rigid lattice system. Furthermore, we predict that the KK interlayer exciton becomes lower in energy than the intralayer KK state due to the strain-induced potentials at low twist angles. The lowest lying momentum-dark K$\Lambda$ exciton is also predicted to undergo a larger red-shift than in the rigid lattice, visible in PL spectra. Most importantly, we find a qualitative change in the absorption spectrum in naturally stacked bilayers reflecting the bright KK excitons, where multiple peaks emerge as a consequence of the strain-induced potentials - in stark contrast to the single peak that is present in the rigid lattice. Consequently, we predict an unambiguous signature of atomic reconstruction in the optical response of moir\'e excitons in naturally stacked twisted TMD homobilayers.

\section{Moir\'e potential in atomically reconstructed TMD bilayers}
To gain access to the moir\'e exciton energy landscape in atomically reconstructed bilayers, we first describe the change in geometry in the superlattice. Following the approach in Refs.\cite{enaldiev2020stacking,carr2018relaxation} we set up a functional for the stacking energy, taking into account the strain energy from continuum theory and a parameterized equation for the adhesion energy between the layers, where the model parameters have been fitted to data from density functional theory in Refs.\cite{enaldiev2020stacking,ferreira2021band}. The functional can then be turned into an optimization problem by expanding the displacement vectors $\mathbf{u}^{l}(\mathbf{r})$ as a Fourier expansion $\mathbf{u}^{l}(\mathbf{r})=\sum_n \mathbf{u}^{l}_ne^{i\mathbf{g}_n\mathbf{r}}$ and minimizing the integral with respect to the Fourier coefficients $\mathbf{u}^{l}_n$. Here, $l$ is the layer index, $\mathbf{\mathbf{g}}_n$ are the reciprocal vectors of the moir\'e lattice and $\mathbf{r}$ is the real space coordinate in the supercell. For further details, see the supplemental part (SI) of the paper \cite{supp} (see also references \cite{duerloo2012intrinsic,rostami2018piezoelectricity,bernardini1997spontaneous,cappelluti2013tight,rostami2015theory,PhysRevB.99.125424,katsch2018theory,laturia2018dielectric,landau1986theory} therein).

\begin{figure}[t!]
\hspace*{-0.5cm}  
\includegraphics[width=1.00\columnwidth]{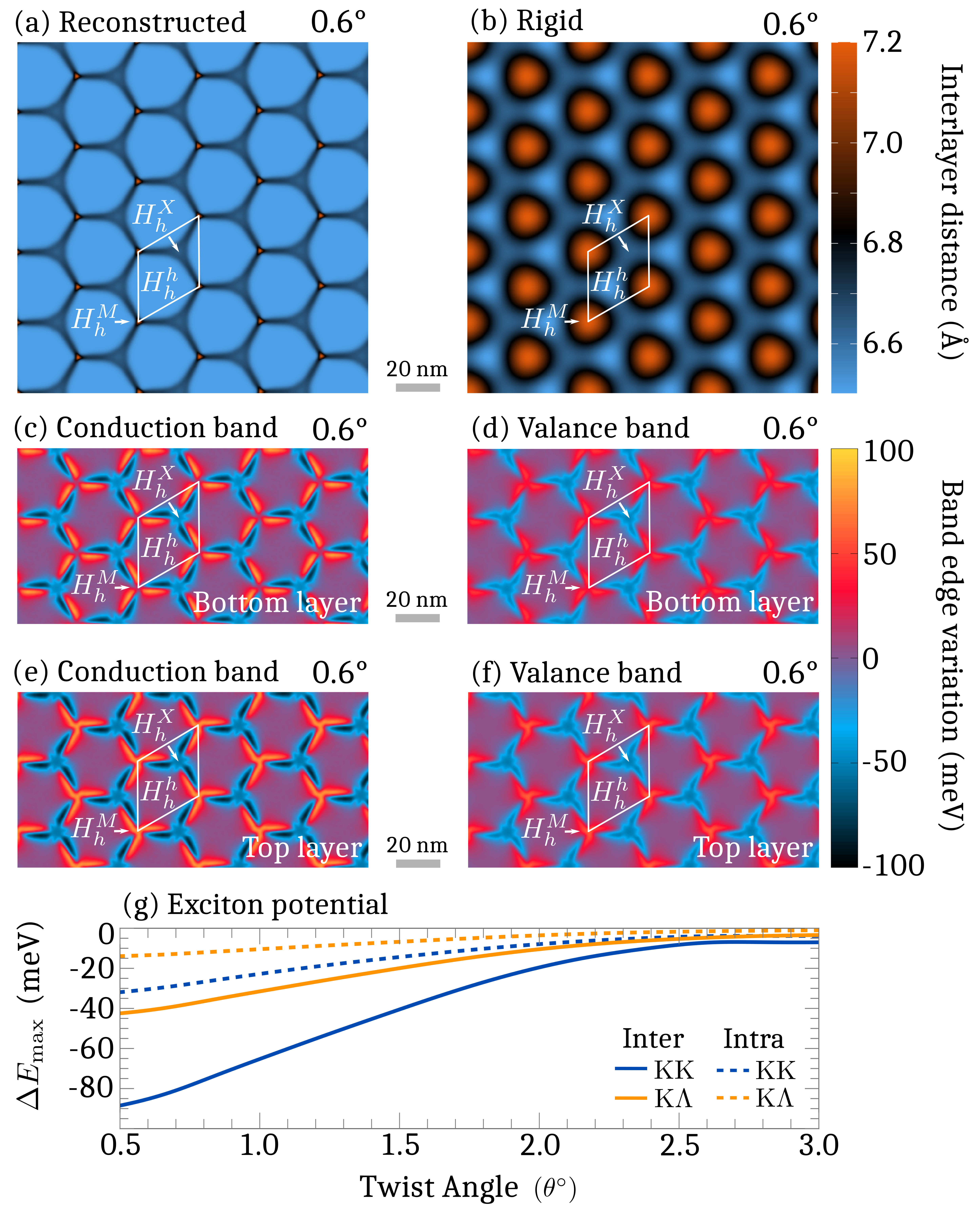}
\caption{\label{fig:1} Spatial map of the interlayer distance in \textbf{(a)} a reconstructed lattice and \textbf{(b)} a rigid lattice at $\theta=0.6^{\circ}$ for H-type stacked WSe$_2$ homobilayers. Here, the interlayer distance is associated with a high-symmetry stacking. The band edge variation for the conduction band in \textbf{(c)} the bottom layer and \textbf{(e)} the top layer at the K-point, and the valance band edge variation in \textbf{(d)} the bottom layer and \textbf{(f)} the top layer at the K-point. Note the reversed sign between the layers and the difference in the potential depth between valance and conduction band. Since this is a purely strain-induced potential, it is not present in the rigid lattice. The shape of the potential stems from the linear combination of the scalar strain potential and the piezo potential. \textbf{(g)} Maximum exciton potential depth $\Delta E_{max}$ in the supercell is shown as a function of twist angle for both intralayer excitons (dashed) and interlayer excitons (solid).}
\end{figure}

After obtaining the displacement vectors, we study the geometrical change in the supercell. Here, we find that the energetically more favorable $H_h^h$ stacking grows into hexagonal domains (Kagome pattern) at small twist angles, separated by a thin boundary wall - in good agreement with experimental studies \cite{weston2020atomic} and previous theoretical works \cite{enaldiev2020stacking}. This reconstruction is illustrated in \autoref{fig:1}a showing the interlayer distance for H-type stacked WSe$_2$ homobilayers at the twist angle $\theta=0.6^{\circ}$. Each high-symmetry stacking is reflected by a specific interlayer distance, which can be computed from DFT \cite{MoireExcitonsLinderalv} and thus the interlayer distance can be used as an indicator for a change in the local atomic registry. The blue domains denote the $H_h^h$ regions, which are significantly larger than in the rigid lattice, cf. \autoref{fig:1}.b. The other high-symmetry stackings ($H_h^X$ and $H^M_h$) consequently shrink in size and form thin boundary networks (cf. \autoref{fig:1}.a). We focus here on the examplary H-type stacked WSe$_2$ homobilayer, but the same approach holds for all vertically stacked TMDs. In R-type stacked homobilayers, we instead obtain large triangular patterns as a result of the atomic reconstruction, which yields a different potential landscape (see SI for further details).

The impact of atomic reconstruction on the exciton energy landscape does not only stem from the geometrical change in the superlattice, but also from the strain that accompanies the lattice deformation. Assuming small displacements, the displacement vector $\mathbf{u}^{l}(\mathbf{r})$ can be associated to the linear strain tensor $\varepsilon_{ij}=\frac{1}{2}(u_{i,j}+u_{j,i})$, where $u_{i,j}=\frac{1}{2}(\partial_iu_j+\partial_ju_i)$ and $i(j)=(x,y)$. Due to its D3h symmetry, we can decompose the dominating strain contributions to the moir\'e potential into two parts (see SI) \cite{amorim2016novel,cazalilla2014quantum,fang2018electronic}: scalar potential composed of inhomogeneous uniaxial strain in both x- and y-direction \cite{khatibi2018impact,enaldiev2022self} and inhomogeneous vector gauge potential known as piezo potential \cite{enaldiev2020stacking,ferreira2021band,Enaldiev_2021}. Details about the strain potential model are provided in the SI.

In addition to the strain-induced potentials, there is a stacking-dependent alignment shift of the two monolayer band structures, originating from charge polarization between the two layers, that already exists in the non-reconstructed system \cite{brem2020tunable,PhysRevResearch.3.043217}. In a homobilayer, these only affect the energy of interlayer excitons \cite{PhysRevResearch.3.043217}. Furthermore, it is only present in R-type stacked bilayers due to the lack of inversion symmetry, which allows for local charge transfers between the layers \cite{tong2020interferences}. The remaining interlayer coupling mechanism contributing to the total moir\'e potential is given by the interlayer hybridization. For some electronic states, in particular in the $\Lambda$ valley in the conduction band, the large wave function overlap between the layers leads to a strong mixing of states in both layers, which can be described as a mixing of interlayer and intralayer excitons into hybrid exciton states \cite{PhysRevResearch.3.043217,D0NR02160A}. In H-type stacking, the strongest hybridization occurs around the $H^h_h$ stacking due its shorter interlayer distance (cf. \autoref{fig:1}.a-b). 
To sum up, we have four different contributions to the moir\'e potential: 1) scalar strain potential $S^{\mathbf{\xi}}_{L\mathbf{g}}$, 2) piezo potential $P^{\mathbf{\xi}}_{L\mathbf{g}}$, 3) stacking-dependent alignment potential $V^{\mathbf{\xi}}_{L\mathbf{g}}$ and 4) tunneling $T^{\mathbf{\xi}}_{LL^{\prime}\mathbf{g}}$. The first three act as a periodic renormalization of exciton energies, i.e. they are diagonal with respect to the layer index, and can thus be modeled as an external potential acting on free excitons $M^{\mathbf{\xi}}_{L\mathbf{g}}=S^{\mathbf{\xi}}_{L\mathbf{g}}+P^{\mathbf{\xi}}_{L\mathbf{g}}+V^{\mathbf{\xi}}_{L\mathbf{g}}$, where $L=(l_e,l_h)$ is a compound layer index and $\mathbf{\xi}=(\xi_e,\xi_h)$ is the exciton valley index. 

In \autoref{fig:1}.c-d, we show the resulting band edge variation for the conduction band (\autoref{fig:1}.c) and the valance band (\autoref{fig:1}.d) in the bottom layer at $\theta=0.6^{\circ}$. Similarly, the band edge variation for the top layer is shown in \autoref{fig:1}.e-f, where the conduction band is given by \autoref{fig:1}.e and the valance band is given by \autoref{fig:1}.e at $\theta=0.6^{\circ}$. Here, the only potentials that are contributing are the scalar strain potential and the piezo potential. In general, homobilayers mostly reconstruct due to atomic rotation \cite{van2023rotational}, which is associated with the piezo potential and will be concentrated at the $H_h^M$/$H_h^X$ sites. However, there is also a significant scalar strain potential close to the boundary regions of the $H_h^h$ domains which also significantly contribute to the band edge variation. The combination of scalar potential and piezo potential in turn gives the tilted star-shape of the total band edge variation. Comparing the variation for different layers (\autoref{fig:1}.e to \autoref{fig:1}.d ) we find that the minima of the conduction band and the maxima of the valance band occur at the same sites, thus yielding a very efficient band gap renormalization for interlayer excitons. In contrast, comparing \autoref{fig:1}.c and \autoref{fig:1}.d we find that the minima for the conduction band and the maxima for the valance band occur at different sites in the supercell. This means that there is a certain driving force for charge separation. However, the binding energies of the intralayer excitons are very strong ($\sim$ -140 meV for KK) which in turn means that one would need a very deep potential in order to have a complete charge separation. At angles $\theta>0.5^{\circ}$, the formation of excitons is still favored in comparison to the separated unbounded electron/holes. For much smaller supercells, however, the formation of in-plane charge transfer excitons can occur as reported in \cite{naik2022intralayer}.

The effective exciton potential as shown in \autoref{fig:1}.g gives the calculated potential depth for KK (blue) and K$\Lambda$ (yellow) excitons as a function of twist angle, both for intra- (dashed) and interlayer (solid) excitons. Since the piezo potential in H-type stacking is the same for both the valence and the conduction band in both layers, the only contributing factor is the scalar strain potential. The color map of this effective potential is given in the supplemental part of the paper. Below $\theta=2^{\circ}$ we find a significant decrease of the KK interlayer exciton potential (solid blue) due to the efficient band gap renormalization stemming from the scalar strain potential. The KK intralayer (dashed blue) potential has a much shallower potential due to the conduction and valance band shifting in the same direction, although with different rates. Comparing to the effective K$\Lambda$ (solid yellow) potential, it only reaches half the potential depth of the KK. This can be understood from different orbital compositions of the valleys, which directly impact the change in the orbital overlap with scalar strain \cite{PhysRevB.96.045425}. Note that these effective exciton potentials are all in stark contrast to the rigid lattice where they do not occur at all for H-type stacking. For R-type stacking, we predict in general much weaker band edge variations, having only a maximum variation of $\sim$40 meV. This difference comes primarily from the lack of scalar strain in R-type stacking, were the domain walls are formed purely from shear strain - in good qualitative agreement with the experimentally observed values \cite{van2023rotational}. In turn, this means that the effective intralayer exciton potential will be close to $0$ meV. There will, however, be a significant addition to the interlayer potential in the form of the piezo potential (see SI).

\section{Moir\'e exciton energy landscape in reconstructed TMD bilayers}
Having a microscopic access to the potential landscape of moir\'e excitons, we can now set up the complete moir\'e exciton Hamilton operator. Here, we start from a decoupled monolayer basis in electron-hole picture and add the moir\'e potential as periodic modifications to these energies \cite{hagel2022electrical,brem2020tunable,D0NR02160A}. Transforming to an exciton basis, the Hamiltonian then reads 
\begin{equation}\label{eq:ExcitonHam}
\begin{split} &H_0=\sum_{L\mathbf{Q}\mathbf{\xi}}E^{\mathbf{\xi}}_{L\mathbf{Q}}X^{\mathbf{\xi}\dagger}_{L\mathbf{Q}}X^{\mathbf{\xi}}_{L\mathbf{Q}}+\sum_{L\mathbf{Q} \mathbf{\xi}\mathbf{g}}M^{\mathbf{\xi}}_{L\mathbf{g}}X^{\mathbf{\xi}\dagger}_{L\mathbf{Q}+\mathbf{g}}X^{\mathbf{\xi}}_{L\mathbf{Q}}\\
    &+\sum_{LL^{\prime}\mathbf{Q}\mathbf{\xi}\mathbf{g}}T^{\mathbf{\xi}}_{LL^{\prime}\mathbf{g}}X^{\mathbf{\xi}\dagger}_{L\mathbf{Q}+\mathbf{g}}X^{\mathbf{\xi}}_{L^{\prime}\mathbf{Q}} +h.c,
    \end{split}
\end{equation}
where $E^{\mathbf{\xi}}_{L\mathbf{Q}}$ is the binding energy of the decoupled monolayer as obtained from the generalized Wannier equation \cite{ovesen2019interlayer} and $\mathbf{Q}$ is the center-of-mass momentum. The effective masses for electrons and holes are obtained from Ref.\cite{Korm_nyos_2015}. Furthermore, $X^{(\dagger)}$ denotes the annihilation (creation) operator of excitons. Moreover, $M^{\mathbf{\xi}}_{L\mathbf{g}}$ is the combined layer-diagonal moir\'e potential component and $T^{\mathbf{\xi}}_{LL^{\prime}\mathbf{g}}$ is the tunneling matrix element. The latter gives rise to the hybridization of intra- and interlayer excitons, which corresponds to the off-diagonal component of the moir\'e potential \cite{MoireExcitonsLinderalv}.

\begin{figure}[t!]
\hspace*{0.0cm} 
\includegraphics[width=1.00\columnwidth]{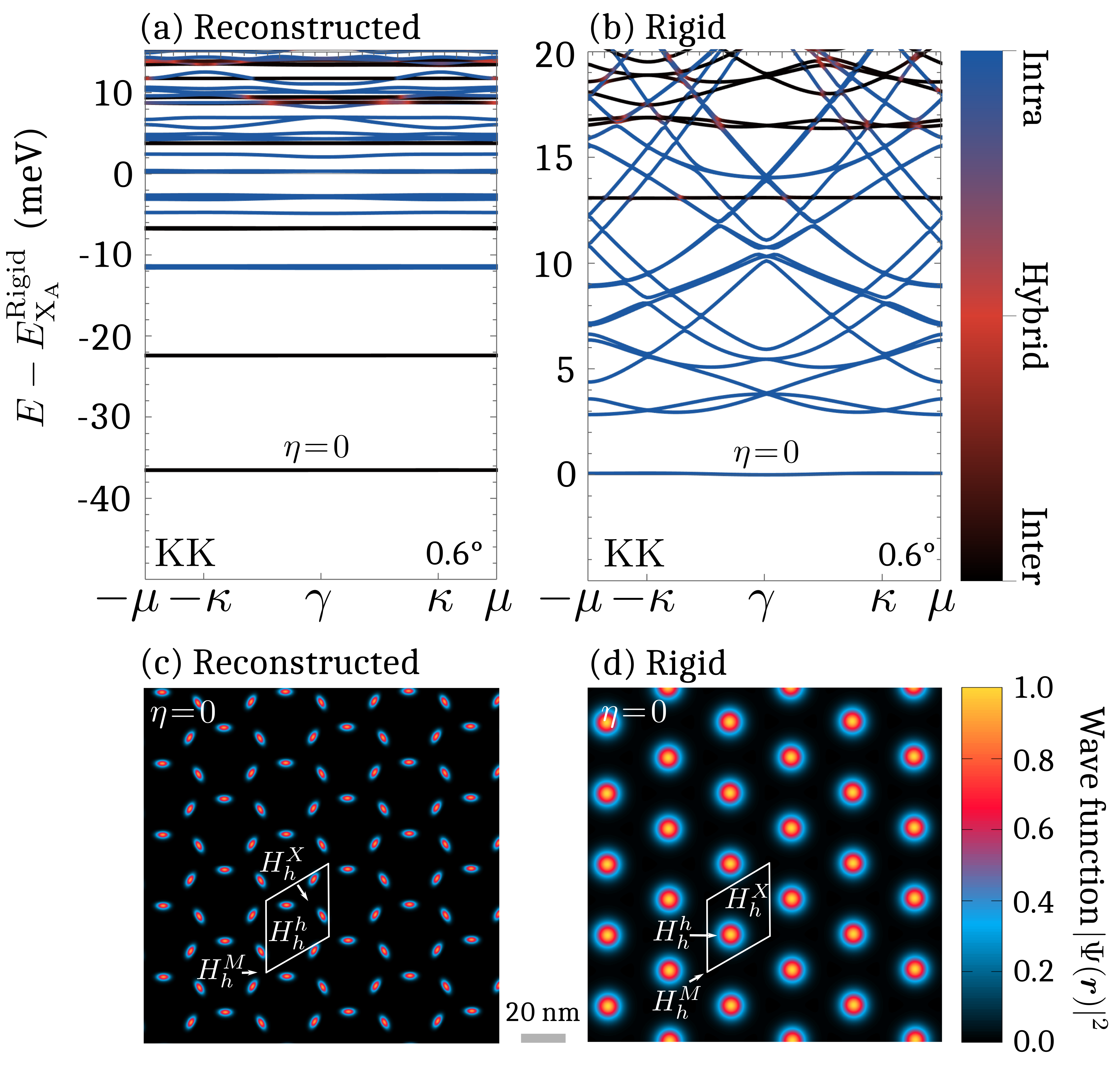}
\caption{\label{fig:2} Band structure and wave function of the bright KK exciton in H-type stacked WSe$_2$ bilayer at $\theta=0.6^{\circ}$ in \textbf{(a,c)} the reconstructed lattice and \textbf{(b,d)} the rigid lattice. Energies are normalized to the lowest-lying KK exciton in the rigid lattice. Note that the reconstruction drastically changes the localization site of excitons and that the efficient moir\'e potential for interlayer excitons shifts them below the intralayer KK states. The band index $\eta=0$ indicates which wave function is shown.}
\end{figure}

The periodicity of the superlattice is taken into account by applying the zone-folding scheme \cite{hagel2022electrical,D0NR02160A,brem2020tunable}, where we restrict the summation over $\mathbf{Q}$ to the first mini-Brillouin zone (mBZ), where states with larger momentum are folded back via mBZ lattice vectors $\mathbf{g}$ giving rise to the formation of mini-subbands. The final moir\'e exciton energies are then obtained by first expanding the exciton operators into a moir\'e exciton basis $Y^{\dagger}_{\mathbf{\xi}\eta\mathbf{Q}}=\sum_{\mathbf{g}L}\mathcal{C}^{\mathbf{\xi}\eta*}_{L\mathbf{g}}(\mathbf{Q})X^{\mathbf{\xi}\dagger}_{L,\mathbf{Q}+\mathbf{g}}$, where $Y^{(\dagger)}$ are the moir\'e exciton annihilation (creation) operators, $\eta$ is the new band index, and $\mathcal{C}^{\mathbf{\xi}\eta*}_{L\mathbf{g}}(\mathbf{Q})$ are the mixing coefficients that mix different sub-bands and layer configurations. Applying this transformation to \autoref{eq:ExcitonHam} we obtain the moir\'e exciton eigenvalue equation
\begin{equation}\label{eq:eigenvalue}
    \begin{split}
        E^{\mathbf{\xi}}_{L\mathbf{Q+g}}\mathcal{C}^{\mathbf{\xi}\eta}_{L\mathbf{g}}(\mathbf{Q})+\sum_{\mathbf{g}^{\prime}}M^{\mathbf{\xi}}_L(\mathbf{g},\mathbf{g}^{\prime})\mathcal{C}^{\mathbf{\xi}\eta}_{L\mathbf{g}^{\prime}}(\mathbf{Q})\\
        +\sum_{L^{\prime}\mathbf{g}^{\prime}}T^{\mathbf{\xi}}_{LL^{\prime}}(\mathbf{g},\mathbf{g}^{\prime})\mathcal{C}^{\mathbf{\xi}\eta}_{L^{\prime}\mathbf{g}^{\prime}}(\mathbf{Q})=\mathcal{E}^{\mathbf{\xi}}_{\eta\mathbf{Q}}\mathcal{C}^{\mathbf{\xi}\eta}_{L\mathbf{g}}(\mathbf{Q}).
    \end{split}   
\end{equation}
Here, $\mathcal{E}^{\mathbf{\xi}}_{\eta\mathbf{Q}}$ are the final moir\'e exciton energies, which can be obtained by solving \autoref{eq:eigenvalue} numerically. 

In \autoref{fig:2}, we show the calculated moir\'e exciton band structure for H-type stacked hBN-encapsulated WSe$_2$ bilayers (for R-type stacking see SI) with a reconstructed lattice (\autoref{fig:2}.a) and with a rigid lattice (\autoref{fig:2}.b). We find a large red-shift in the reconstructed lattice, where the lowest sub-band is located $\sim$36 meV below the rigid lattice energies. Interestingly, due to the efficient moir\'e potential for the interlayer exciton, the nature of the lowest lying state has changed from an intralayer exciton to an interlayer exciton. While in the rigid lattice (cf. \autoref{fig:2}.b) most bands have a significant curvature, we find multiple ultra-flat bands in the reconstructed lattice (cf. \autoref{fig:2}.a), both for the interlayer excitons (black) and the intralayer excitons (blue). The spacing between the first two bands is about 15 meV, which is significantly larger than the small spacing (of about 3 meV) in the rigid lattice indicating a much stronger spatial confinement. This is a direct consequence of the additional strain potentials in the reconstructed lattice. Note that also the hybridization of the higher-lying bands is strongly suppressed as there is a lack of band crossing due to their flatness in the reconstructed lattice, cf. the color gradient in \autoref{fig:2}.a-b denoting the degree of hybridization. We need to go to higher energies to see some hybridization in the reconstructed case (red color around 10 meV in \autoref{fig:2}.a). 

The moir\'e exciton wave function in real space $|\Psi(\mathbf{r})|^2$ for the lowest-lying states are shown in \autoref{fig:2}c-d. We find that in the reconstructed lattice the wave function is localized close to the edges of the $H_h^h$-domains of the of the moir\'e unit cell (cf. \autoref{fig:2}.c), reflecting the potential minima from the strain-induced potential. In contrast, in the rigid lattice with no additional strain potential (cf. \autoref{fig:2}.d) there is a much larger probability distribution centered at the $H^h_h$ stacking configuration (cf. \autoref{fig:1}.b). Here, the moir\'e potential only consists of the tunneling and therefore the exciton localization is determined only by the stacking with the strongest tunneling \cite{PhysRevResearch.3.043217}. When increasing the twist angle, the reconstructed lattice starts to revert back to the rigid lattice, making the tunneling the dominating contribution of the moir\'e potential. This change between different regimes implies the possibility of tuning the localization of moir\'e excitons, which should have important consequences for exciton transport in these materials \cite{malic2023exciton,wagner2021nonclassical}.

\begin{figure}[t!]
\hspace*{-0.5cm}  
\includegraphics[width=1.00\columnwidth]{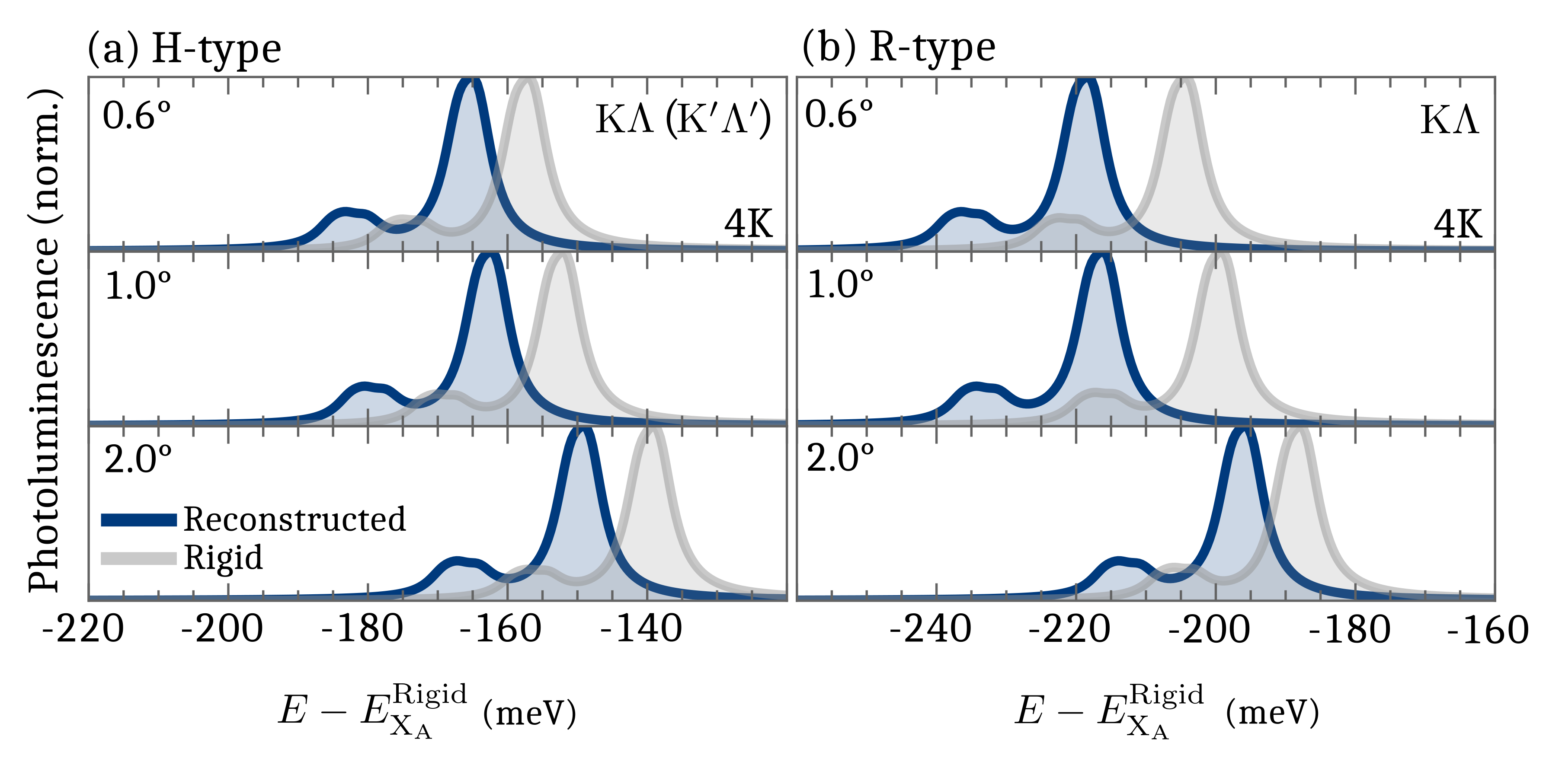}
\caption{\label{fig:3} Normalized photoluminescence spectrum at $4$K at different twist angles, displaying phonon replicas of the energetically lowest K$\Lambda$ exciton in \textbf{(a)} H-type and \textbf{(b)} R-type WSe$_2$ homobilyers. The spectra from the reconstructed and the rigid lattice are shown in blue and gray, respectively. A signature of lattice reconstruction is a more pronounced red-shift of the exciton resonance for smaller twist angles.
}
\end{figure}

\section{Impact of reconstruction on optical spectra}
Having access to the moir\'e exciton energy landscape, we can now determine the optical response of TMD bilayers in presence of atomic reconstruction. In WSe$_2$ homobilayers, the lowest-lying state in R-type stacking is the strongly hybridized K$\Lambda$ exciton, while in H-type stacking K$\Lambda$ and K$^{\prime}\Lambda^{\prime}$ excitons are the lowest ones. Note that the latter two are degenerate in H-type stacking, but they have a reversed layer configuration, as a consequence of the reversed spin-orbit coupling the layers \cite{PhysRevResearch.3.043217,D0NR02160A,tagarelli2023electrical}. We first focus on PL spectra, where these dark states are expected to be visible via indirect phonon sidebands \cite{PLBrem}. Here, the exciton can not directly emit a photon due to the large momentum mismatch between the valleys. Instead, it decays via a two-step process, first scattering to a virtual bright state within the light cone via phonon emission followed by the emission of a photon \cite{PLBrem}. 

Figure \ref{fig:3}.a-b shows the PL spectra for WSe$_2$ with H-type (cf. \autoref{fig:3}.a) and R-type stacking (cf. \autoref{fig:3}.b) at 4K at different twist angles. The reference frame for the energies are the same as in \autoref{fig:2}.a-b. We observe sidebands from momentum-dark states, where the smaller (larger) peak stems from optical (acoustic) phonon modes \cite{D0NR02160A,PhysRevResearch.3.043217}. The energetic position of the K$\Lambda$(K$^{\prime}\Lambda^{\prime}$) phonon replicas will primarily depend on the interlayer hybridization, stemming from the strong interlayer tunneling of electrons around the $\Lambda$($\Lambda^{\prime}$) valley. This gives rise to a twist-angle-dependent dehybridization, in turn yielding a blue shift when increasing the twist angle (gray peaks in \autoref{fig:3}), i.e. the larger the twist-angle the smaller is the wave function overlap \cite{D0NR02160A,merkl2020twist}. Since this component of the moir\'e potential is already present in the rigid lattice we find a very similar behavior in both the rigid and the reconstructed lattice when changing the twist angle. There is, however, a clearly larger red-shift for the reconstructed lattice at lower twist angles due to the increasing strain-induced potentials. This is especially noticeable for R-type stacking where the piezo potential contributes to the effective exciton potential. Consequently, the change in the peak position closely resembles the change in the effective exciton potential for K$\Lambda$ (SI Fig.3). In \autoref{fig:3}.b we can also see that the peak position of the reconstructed lattice is converging towards the peak position of the rigid lattice when increasing the twist angle, indicating that the lattice becomes more rigid at larger angles.

\hfill
\begin{figure}[t!]
\hspace*{-0.5cm}  
\includegraphics[width=1.0\columnwidth]{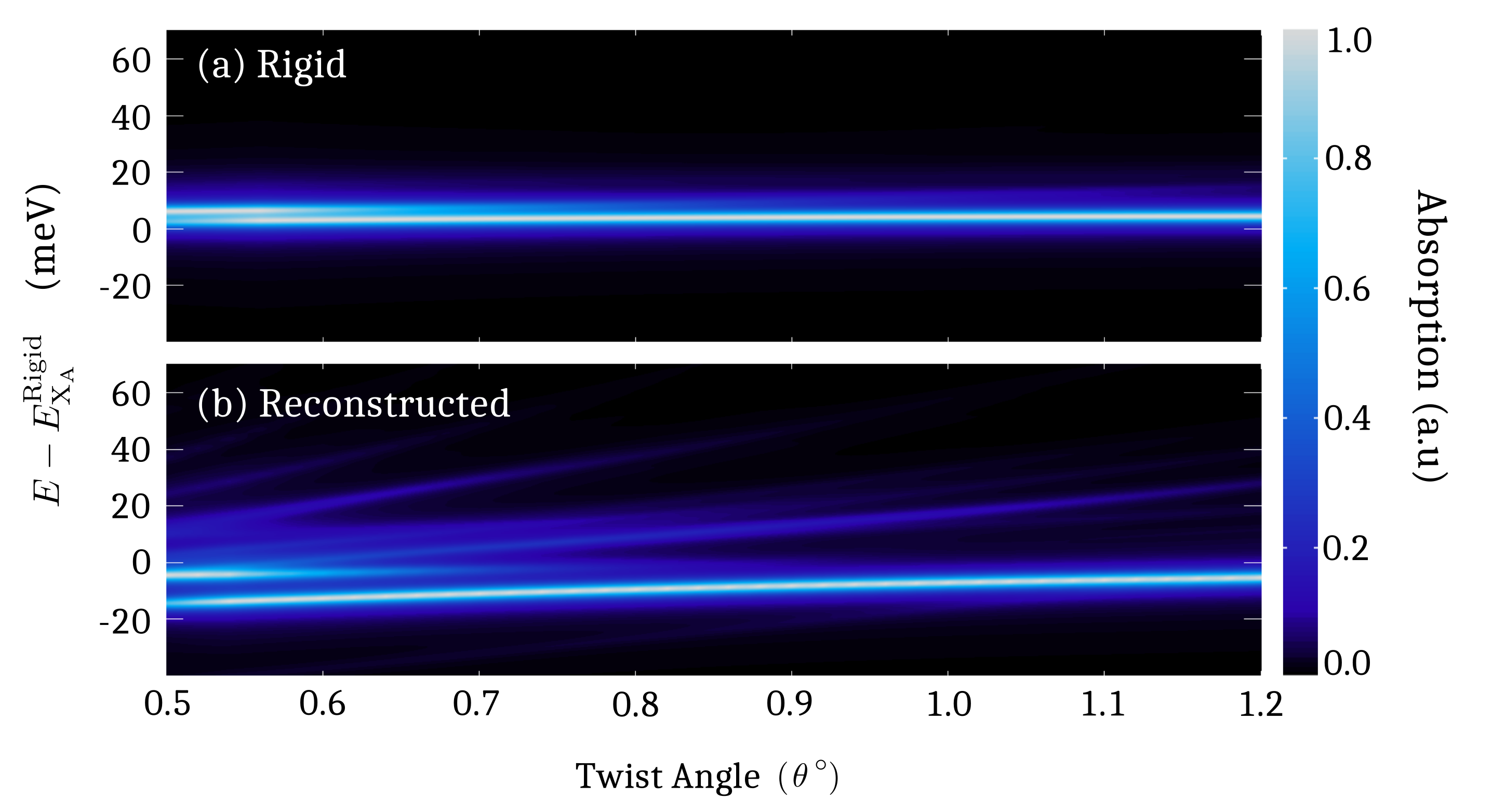}
\caption{\label{fig:4} Twist-angle dependence of the normalized absorption in H-type WSe$_2$ bilayers for \textbf{(a)} the rigid lattice and \textbf{(b)} the reconstructed lattice. The emergence of multiple peaks in the latter is due to the strain-induced intralayer potential and is an unambiguous signature of lattice reconstruction.}
\end{figure}

So far we have been focusing on photoluminescence that is dominated by phonon sidebands of the energetically lowest dark K$\Lambda$ excitons. These states are strongly hybridized \cite{D0NR02160A,PhysRevResearch.3.043217}, and are thus subject to a pronounced moir\'e potential already in the rigid lattice. The situation is qualitatively different for bright KK excitons that are accessible in absorption spectra. Here, the reconstructed lattice gives rise to a deep strain-induced potential, whereas the rigid lattice only experiences a small moir\'e potential due to the weak periodic tunneling strength around the K-valley. In \autoref{fig:2}.a we have already demonstrated the significant change occurring in the band structure of KK excitons, where multiple flat-bands have emerged in the reconstructed lattice for both the intralayer and the interlayer excitons.

In \autoref{fig:4}.a-b, we have calculated the absorption spectrum of H-type stacked WSe$_2$ bilayers as a continuous function of the twist angle for both the rigid lattice (\autoref{fig:4}.a) and the reconstructed lattice (\autoref{fig:4}.b). Here, we immediately observe a drastic change in the spectrum, where below $\theta\approx1.0^{\circ}$ multiple peaks appear in a reconstructed lattice - in stark contrast to the single-peak spectra in the rigid lattice that are mostly angle-independent. Furthermore, at smaller angles we find a noticeable splitting around $\sim$15 meV between the two brightest peaks in the reconstructed lattice - in good agreement with experimental observations \cite{andersen2021excitons}. We also find a small splitting in the rigid lattice due to the tunneling of holes, which consequently traps the lowest state at $H_h^h$ (cf. \autoref{fig:2}.d). This state, however, is very close to the next non-trapped state (cf. \autoref{fig:2}.b), and thus the trapping is not very efficient. In the reconstructed lattice (cf. \autoref{fig:4}.b), we instead have a rich optical response, consisting of multiple peaks. Here, the lowest branches stem from excitons trapped in the deep strain potential pockets (cf. \autoref{fig:2}.a/c). Consequently, these peaks undergo a non-linear blue-shift as we increase the twist angle, following the non-linear twist angle dependence of the strain-induced moir\'e potential (cf. \autoref{fig:1}.g). 
Furthermore, in \autoref{fig:2}.a, we find that the lowest lying KK states are of pure interlayer character (black lines in \autoref{fig:2}.a). These are not visible in absorption spectra due to their by orders of magnitude smaller overlap between the electron and the hole wave function compared to intralayer excitons \cite{Merkl2019}. When increasing the twist angle, the number of peaks is reduced until it converges to the rigid lattice case around $\theta=1.2^{\circ}$ that is characterized by only one resonance.

In R-type stacking (shown in the SI), we do not observe the same effect. Here, the strain-induced potential for intralayer KK excitons is negligible and the impact on the spectra only stems from the weak tunneling of carriers. We predict this drastic change in optical absorption due to atomic reconstruction to be similar for all naturally stacked TMD homobilayers. Note that the multi-peak structure is also well-known from TMD heterostructures, such as MoSe$_2$-WSe$_2$, already in the rigid lattice case due to deep moir\'e potentials \cite{brem2020tunable}. Here, the impact of lattice reconstruction is only of quantitative nature with changes for the peak separation and amplitude. The situation is qualitatively different in naturally stacked TMD homobilayers, where the strain-induced potential in the reconstructed lattice is the dominant mechanism and leads to the characteristic multi-peak structure for the bright A exciton. The latter can thus be considered as an unambiguous optical signature for the presence of lattice reconstruction. 

\section{Conclusion}
Overall, our work provides new microscopic insights into many-particle mechanisms dominating the optical response in twisted TMD homobilayers. The atomic reconstruction has a large impact on the moir\'e exciton energy landscape where strain-induced potentials lead to significant red-shifts and multiple new flat bands. Furthermore, we have calculated moir\'e exciton wave functions and predict that excitons are trapped at different sites in the reconstructed lattice reflecting the maxima of the strain potentials. 
Finally, we find a significant change in optical spectra of naturally stacked TMD homobilayers due to reconstruction. In particular, we predict the emergence of a multi-peak structure in absorption spectra - in stark contrast to the single-peak dominated spectra in a rigid lattice. Our work can help guide future experiments in the growing field of 2D material twistronics.

\section{Acknowledgments}
We thank Maja Feierabend for insightful discussions. This project has received funding from Deutsche Forschungsgemeinschaft via CRC 1083 (project B09) and the regular project 512604469.

\end{document}


\preprint{APS/123-QED}

\title{Supplemental information: Impact of atomic reconstruction on optical spectra of twisted TMD homobilayers}

\author{Joakim Hagel}
  \email{joakim.hagel@chalmers.se}
  \affiliation{%
Department of Physics, Chalmers University of Technology, 412 96 Gothenburg, Sweden\\
}%
\author{Samuel Brem}%
\affiliation{%
 Department of Physics, Philipps University of Marburg, 35037 Marburg, Germany\\
}%
\author{Johannes Abelardo Pineiro}
  \affiliation{%
Department of Physics, Philipps University of Marburg, 35037 Marburg, Germany\\
}%
  \author{Ermin Malic}%
  \affiliation{%
 Department of Physics, Philipps University of Marburg, 35037 Marburg, Germany\\
}%
\affiliation{%
Department of Physics, Chalmers University of Technology, 412 96 Gothenburg, Sweden\\
}%

\maketitle
\section{Atomic reconstruction in H-type stacking}
In the main part of the paper the total potential landscape and its corresponding KK moir\'e exciton band structure is discussed for H-type stacking. Here, we show strain fields for H-type stacking and the effective moir\'e exciton potential stemming from the strain. In addition, we also discuss the corresponding K$\Lambda$ band structure.

\subsection{Strain field}

\begin{figure}[b!]
\hspace*{0.0cm}  
\includegraphics[width=1.00\columnwidth]{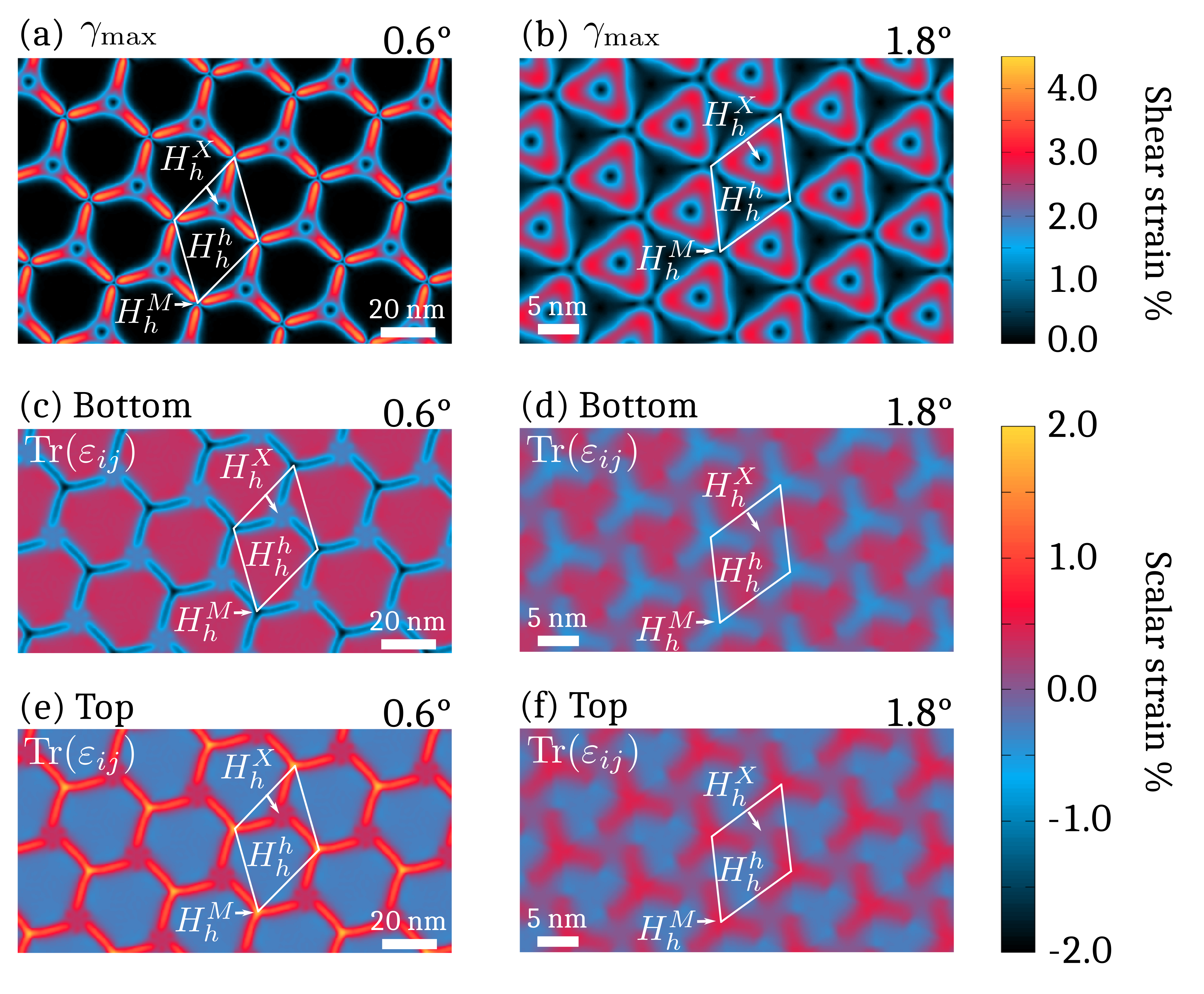}
\caption{\label{fig:strain_supp}Shear strain profile at \textbf{(a)} $\theta=0.6^{\circ}$ and \textbf{(b)} $\theta=1.8^{\circ}$. \textbf{(c-f)} Scalar strain profile for both top and bottom layer at $\theta=0.6^{\circ}$ and $\theta=1.8^{\circ}$.}
\end{figure}

In \autoref{fig:strain_supp}.a-b we show the calculated strain field for the shear strain in the form of the maximum shear strain value $\gamma_{\text{max}}=u_{\text{max}}-u_{\text{min}}$, where $u_{\text{max,min}}$ is given by
\begin{equation}
    u_{\text{max,min}}=\frac{\text{Tr}(\varepsilon_{ij})}{2}\pm\sqrt{\frac{(\varepsilon_{11}-\varepsilon_{22})^2}{2}+\varepsilon_{12}\varepsilon_{21}}.
\end{equation}
Here, $\varepsilon_{ij}$ is the linear strain tensor (see \autoref{sec:strain} for further details). At lower angles (cf. \autoref{fig:strain_supp}.a) we can see that the magnitude of the shear strain increases while the $H_h^h$ region also grows in size. At larger twist angles (cf. \autoref{fig:strain_supp}.b) we can see that the magnitude of strain decreases and instead forms a more triangular pattern around $H^X_h$, in very good agreement with previous works \cite{van2023rotational}. Furthermore, in \autoref{fig:strain_supp}.c-f we show the calculated scalar strain for different angles in both the top layer and the button layer. Here, we predict that at smaller angles (cf. \autoref{fig:strain_supp}.c/e) a pronounced scalar strain will be present along the edges of the $H^h_h$ domains with a range of $\pm$ 2\%. Note that the shear component of the strain is still the dominate component, but that the scalar strain adds a significant contribution to strain profile. At larger angles (cf. \autoref{fig:strain_supp}.d/f) we find that the scalar strain decreases in magnitude to range of approximately $\pm$ 0.4\%, in good agreement with previous theoretical predictions \cite{van2023rotational}.

\subsection{Effective moir\'e exciton potential}
\begin{figure}[t!]
\hspace*{0.0cm}  
\includegraphics[width=1.00\columnwidth]{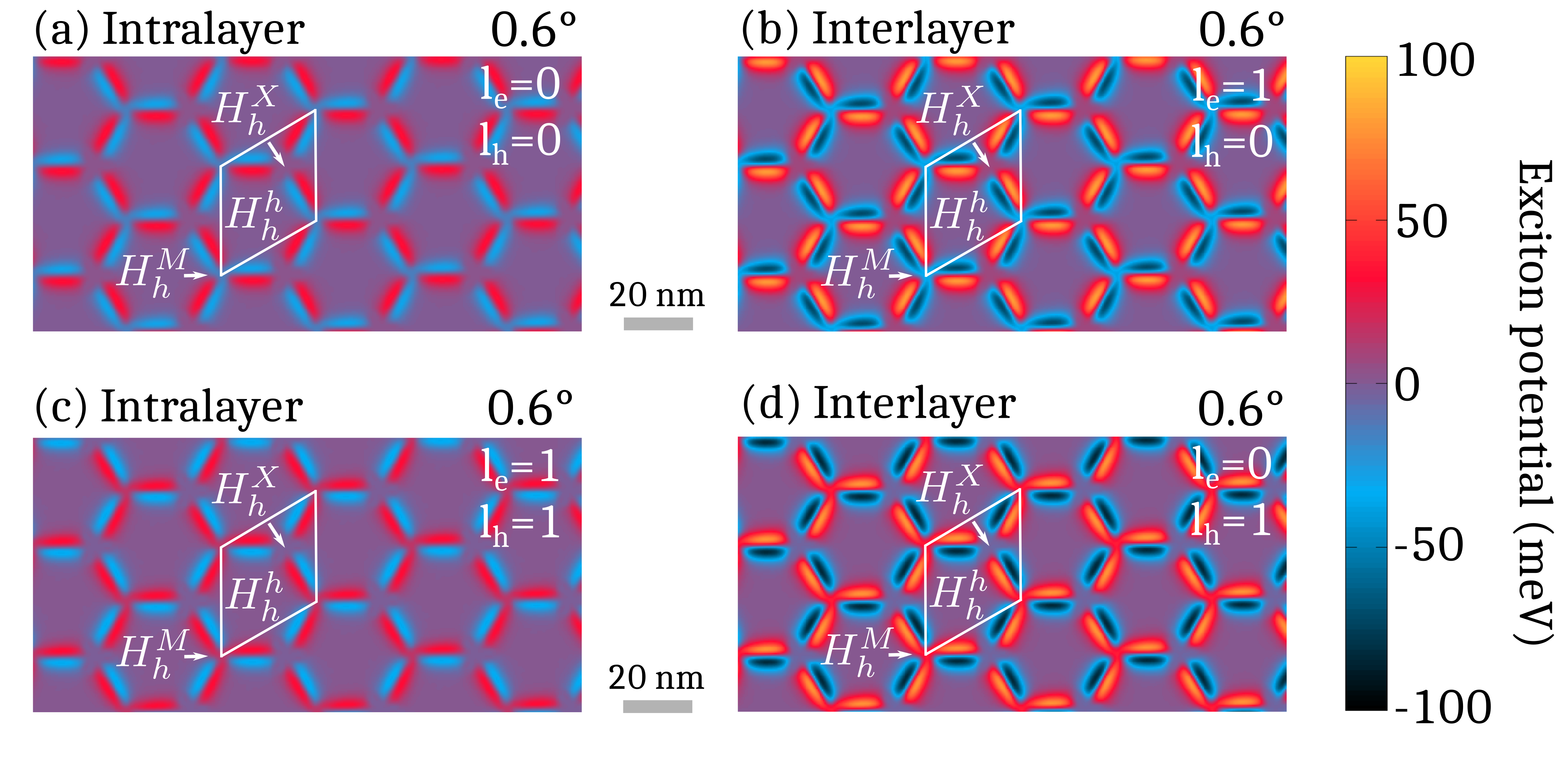}
\caption{\label{fig:1_supp} Effective KK moir\'e exciton potential depth at $\theta=0.6^{\circ}$ for the intralayer exciton in bottom \textbf{(a)} and top \textbf{(c)} layer. The effective moir\'e exciton potential for interlayer excitons is given in \textbf{(b)} for the electron in the top layer and \textbf{(d)} for the electron in the bottom layer. Note that the sign is reversed when changing the layer configuration and that the interlayer exciton potential is deeper than the intralayer one.}
\end{figure}

The total band edge variation shown in Fig.1.c-f in the main part of the paper is the result from the combined scalar strain and piezo potential. In \autoref{fig:1_supp}.a-d we show the effective exciton potential that arises from these band edge variations for the KK exciton. Here, the only contributing component is the scalar strain since the piezo potential is a local rigid shift of the entire band structure. First thing to note is that the potential minimum is localized around the edges of the $H^h_h$ domains with varying sign, stemming from the strong scalar strain close to the edges. Furthermore, the interlayer exciton potential \autoref{fig:1_supp}.b/d has a much deeper potential than the intralayer exciton potential \autoref{fig:1_supp}.a/c. This comes as a consequence of the scalar strain shifting the valance and conduction band in opposite direction for the different layers due to one layer exhibiting compressive strain and the other tensile strain. In turn, this yields a very efficient moir\'e potential for the interlayer excitons which shifts them below the intralayer excitons in the case of KK excitons (Fig.2 in main part). The intralayer excitons also exhibits a significant moir\'e potential, although not as deep as the interlayer potential. This stems from the difference in strain gauge factors for the bands as can be seen from \autoref{tab:gauge}.

\subsection{K$\Lambda$ exciton band structure}

\begin{figure}[t!]
\hspace*{0.0cm}  
\includegraphics[width=1.00\columnwidth]{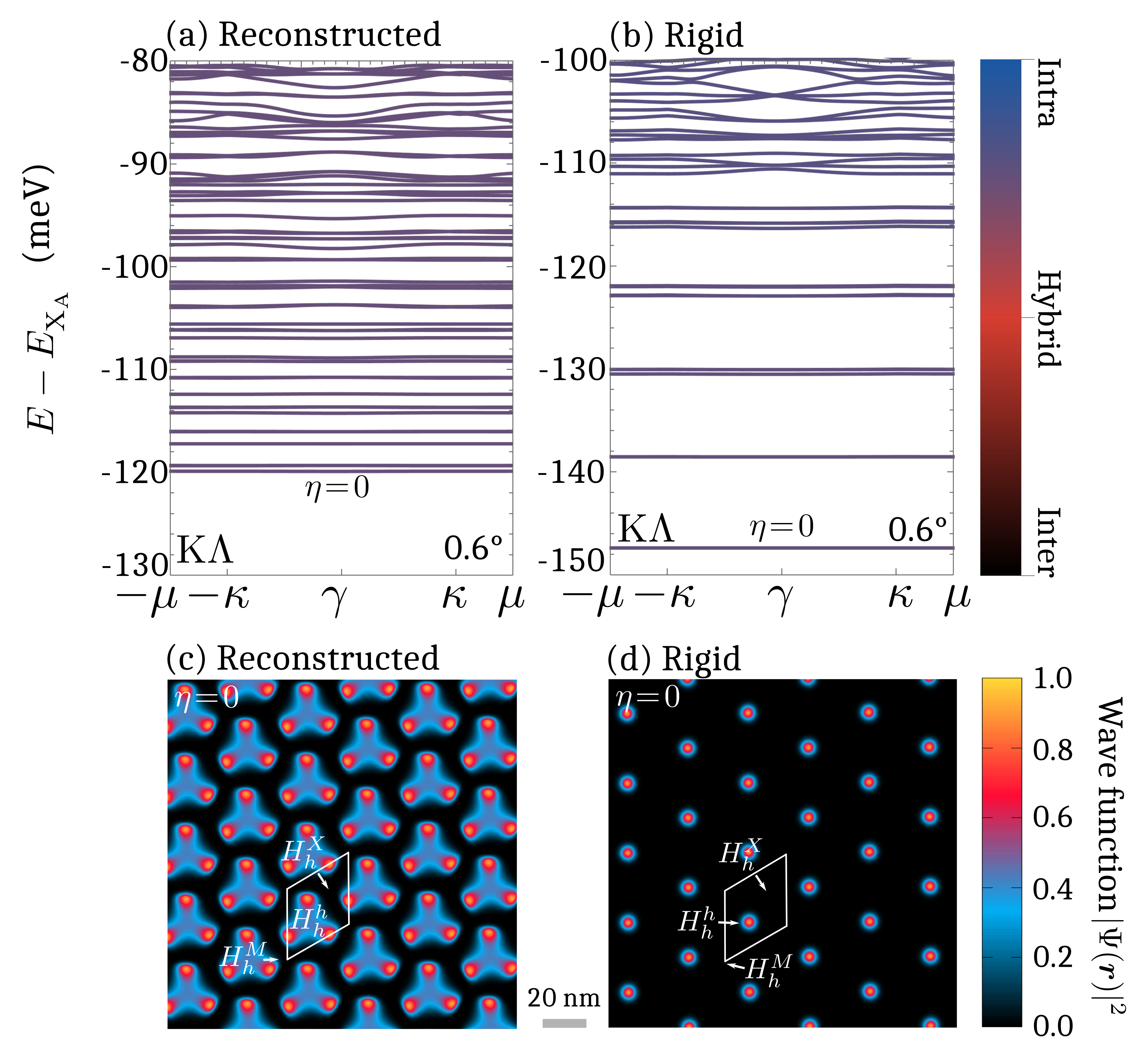}
\caption{\label{fig:2_supp} Band structure for the low-lying dark K$\Lambda$ exciton in H-type stacked WSe$_2$ at $\theta=0.6^{\circ}$ in \textbf{(a)} the reconstructed lattice and \textbf{(b)} the rigid lattice. Energies are normalized to the lowest lying A exciton in their respective lattice. The corresponding K$\Lambda$ exciton wave function for the lowest lying state is shown for \textbf{(c)} the reconstruction lattice and \textbf{(d)} the rigid lattice.}
\end{figure}

In \autoref{fig:2_supp}.a-b we show the calculated moir\'e exciton band structure for the low-lying K$\Lambda$ exciton in H-type stacked WSe$_2$ in the reconstructed lattice (cf. \autoref{fig:2_supp}.a) and the rigid lattice (cf. \autoref{fig:2_supp}.b). Note that the reference point for the energies is given with respect to the A exciton in the reconstructed and the rigid lattice, respectively. Due to the reversed spin-orbit coupling in one of the layers, the energetic distance between the interlayer  and the intralayer K$\Lambda$ exciton is increased. This consequently makes the K$\Lambda$ band structure in H-type stacking less hybridized than in R-type (cf. \autoref{fig:7_supp}.a-b). However, the tunneling still remains the dominant component of the moir\'e potential as can be perceived from the color-gradient of the bands. 

When considering the reconstructed lattice we also observe a decreased distance between the K$\Lambda$ exciton and its respective KK exciton compared to the rigid lattice. This comes as a consequence of the valley dependence of the scalar strain potential that was discussed in Fig. 1.g in the main part of the manuscript. In \autoref{fig:2_supp}.c we show the K$\Lambda$ exciton wave function for the reconstructed lattice, which is, in contrast to the rigid lattice, more spread out and the probability distribution is pushed towards the edges of the $H^h_h$ domains. This comes as a consequence of the competing potentials between electron tunneling and scalar strain, where the most efficient tunneling occurs at the $H_h^h$ sites and the scalar strain is localized at the boundary of this domains. In the rigid lattice, the additional strain potential is not present, which means that K$\Lambda$ exciton wave function is only determined by the tunneling strength, thus yielding a strong localization at the $H^h_h$ site.

\section{Atomic reconstruction in R-type stacking}

\begin{figure}[t!]
\hspace*{-0.5cm}  
\includegraphics[width=1.00\columnwidth]{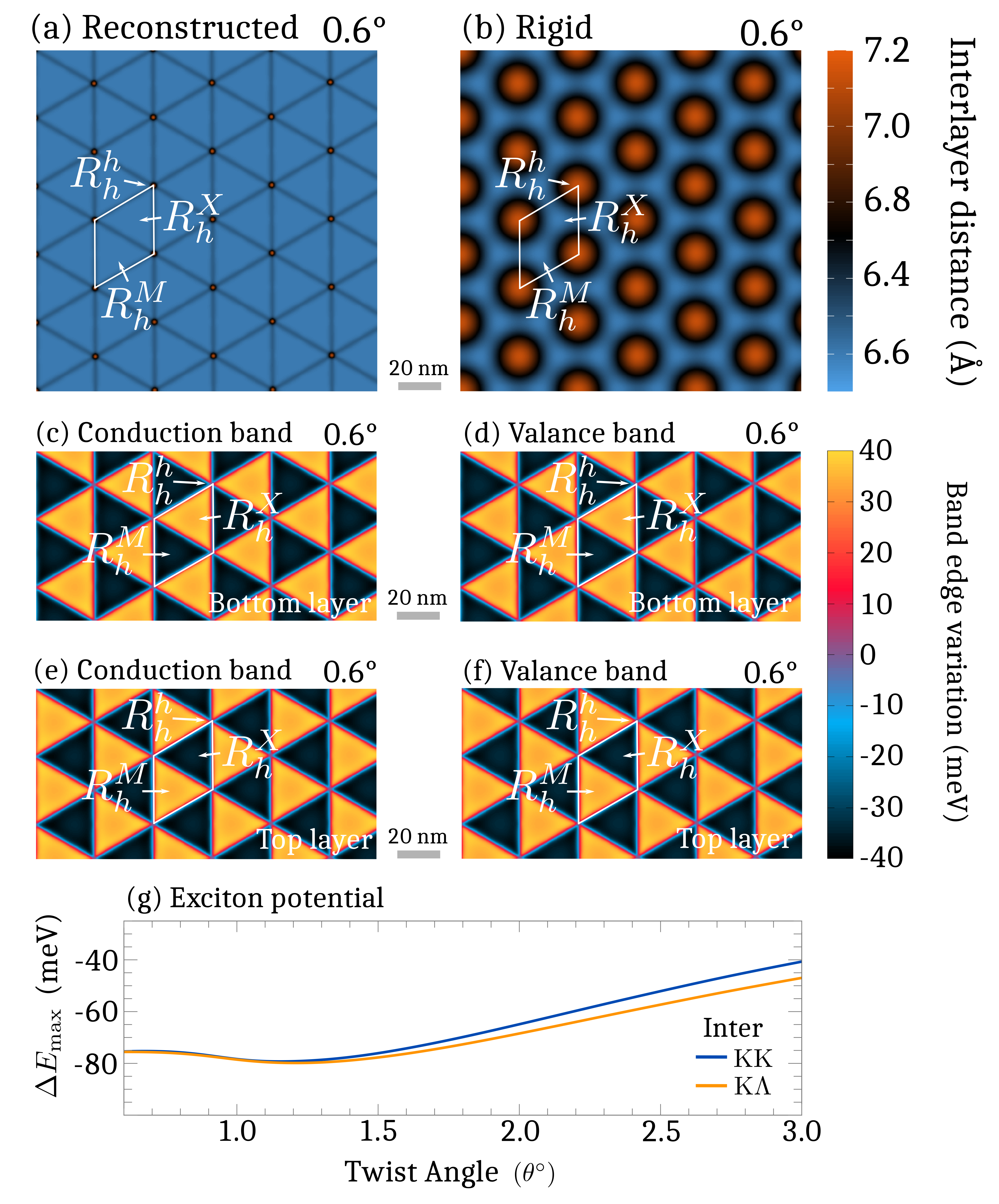}
\caption{\label{fig:3_supp} Spatial map of the interlayer distance in \textbf{(a)} a reconstructed lattice and \textbf{(b)} a rigid lattice at $\theta=0.6^{\circ}$ for R-type stacked WSe$_2$ homobilayers. Here, the interlayer distance is associated with a high-symmetry stacking. The band edge variation for the conduction band in \textbf{(c)} the bottom layer and \textbf{(e)} the top layer at the K-point, and the valance band edge variation in \textbf{(d)} the bottom layer and \textbf{(f)} the top layer at the K-point. Note the reversed sign between the layers. The shape of the potential stems from the linear combination of the the piezo potential and the alignment shift. \textbf{(g)} Maximum exciton potential depth $\Delta E_{max}$ in the supercell is shown as a function of twist angle for interlayer excitons. }
\end{figure}

In this section, we cover the impact of atomic reconstruction on the exciton energy landscape and its optical properties in R-type stacked WSe$_2$. Similarly to Fig.1 in the main part of the paper showing the case of H-type stacking, here we study the geometrical change in the supercell by considering the interlayer distance \autoref{fig:3_supp}.a-b. In contrast to H-type stacking, which exhibits a hexagonal pattern, here we instead have large triangular domains forming. Since the $R^M_h$ and $R^X_h$ stacking are both equivalent in terms of stacking energy, these regions will grow to be the same size. Meanwhile, the less energetically favorable $R^h_h$ stacking will drastically shrink in size. This is in stark contrast to the rigid lattice (cf. \autoref{fig:3_supp}.b) where it is clearly visible. 

The resulting band edge variation for the conduction band is shown in \autoref{fig:3_supp}.c for the bottom layer and \autoref{fig:3_supp}.e for the top layer. Similarly, the variation for the valance band is shown in \autoref{fig:3_supp}.d for the bottom layer and \autoref{fig:3_supp}.f for the top layer. Here, we have two contributing potentials. The first is the alignment shift that is already present in the rigid lattice, but now with a triangular geometry due to the change in local stacking configuration. The second is the piezo potential that stems from the shear strain induced by atomic reconstruction. Note that the scalar potential present in H-type stacking does not appear in R-type stacking for homobilayers. This comes as a consequence from the domain wall formation, which is nearly purely formed from shear strain \cite{enaldiev2020stacking}, in good agreement with experimental observations \cite{van2023rotational}. Comparing \autoref{fig:3_supp}.c with \autoref{fig:3_supp}.d we see that the conduction and valance band is shifting in the same direction in the same layer. Since there is no scalar potential present, the bands also shift the same amount, making for a vanishing moir\'e potential for intralayer excitons. In contrast, by comparing \autoref{fig:3_supp}.c with \autoref{fig:3_supp}.f we find that the conduction band is shifted down where the valance band is shifted up, resulting in a very efficient moir\'e potential in the case of interlayer excitons. This is already expected in the rigid lattice. Here, we have, however, a significant contribution to the depth from the piezo potential, which stands in contrast to H-type stacking, where the piezo potential is the same for both layers, making it negligible when considering the exciton potential.

\begin{figure}[t!]
\hspace*{-0.5cm}  
\includegraphics[width=1.00\columnwidth]{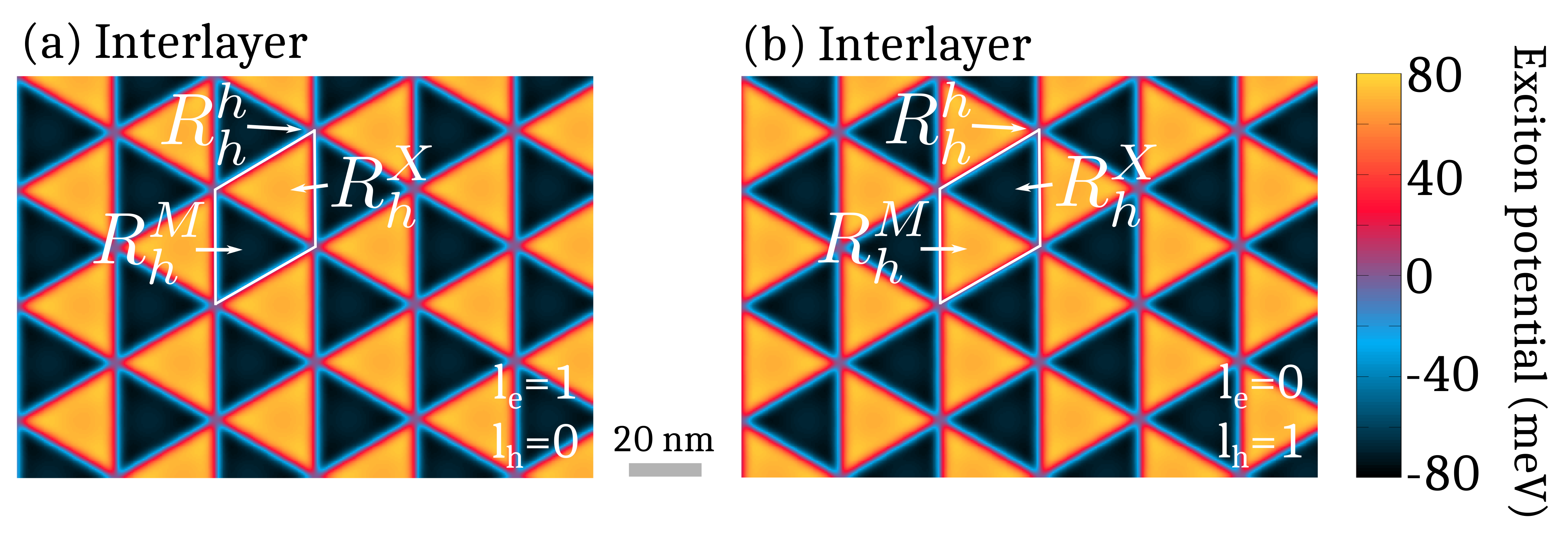}
\caption{\label{fig:4_supp} Effective moir\'e exciton potential depth at $\theta=0.6^{\circ}$ for the KK interlayer exciton for the electron in \textbf{(a)} the top layer and \textbf{(b)} for the electron in the bottom layer. Note that the sign is reversed when changing the layer configuration.}
\end{figure}

In\autoref{fig:3_supp}.g the effective exciton potential depth is shown for interlayer KK and interlayer K$\Lambda$ excitons as a function of the twist angle. Here, we find that the KK and K$\Lambda$ states have the same dependence on $\theta$ up until larger angles where the size of the mBZ increases such that the exciton wave function overlap starts to suppress the potentials, thus yielding a difference between KK and K$\Lambda$ at larger angles. The potential minimum is around $\theta=1.2^{\circ}$ and then it starts to increase again due to the increased size of the supercell - in good agreement with previous theoretical predictions \cite{ferreira2021band}. The spatial map of the KK interlayer potential is shown in \autoref{fig:4_supp}.a-b. Here, we can clearly see the formation of triangular potential wells, stemming from the alignment shift and piezo potential, and consequently trapping interlayer excitons in real space. Furthermore, the sign of the potential between the different layer configurations is different, but with the same depth. Consequently, this means that the interlayer excitons are degenerate, but trapped in different regions.

\subsection{Moir\'e exciton energy landscape}
\begin{figure}[t!]
\hspace*{0.0cm}  
\includegraphics[width=1.00\columnwidth]{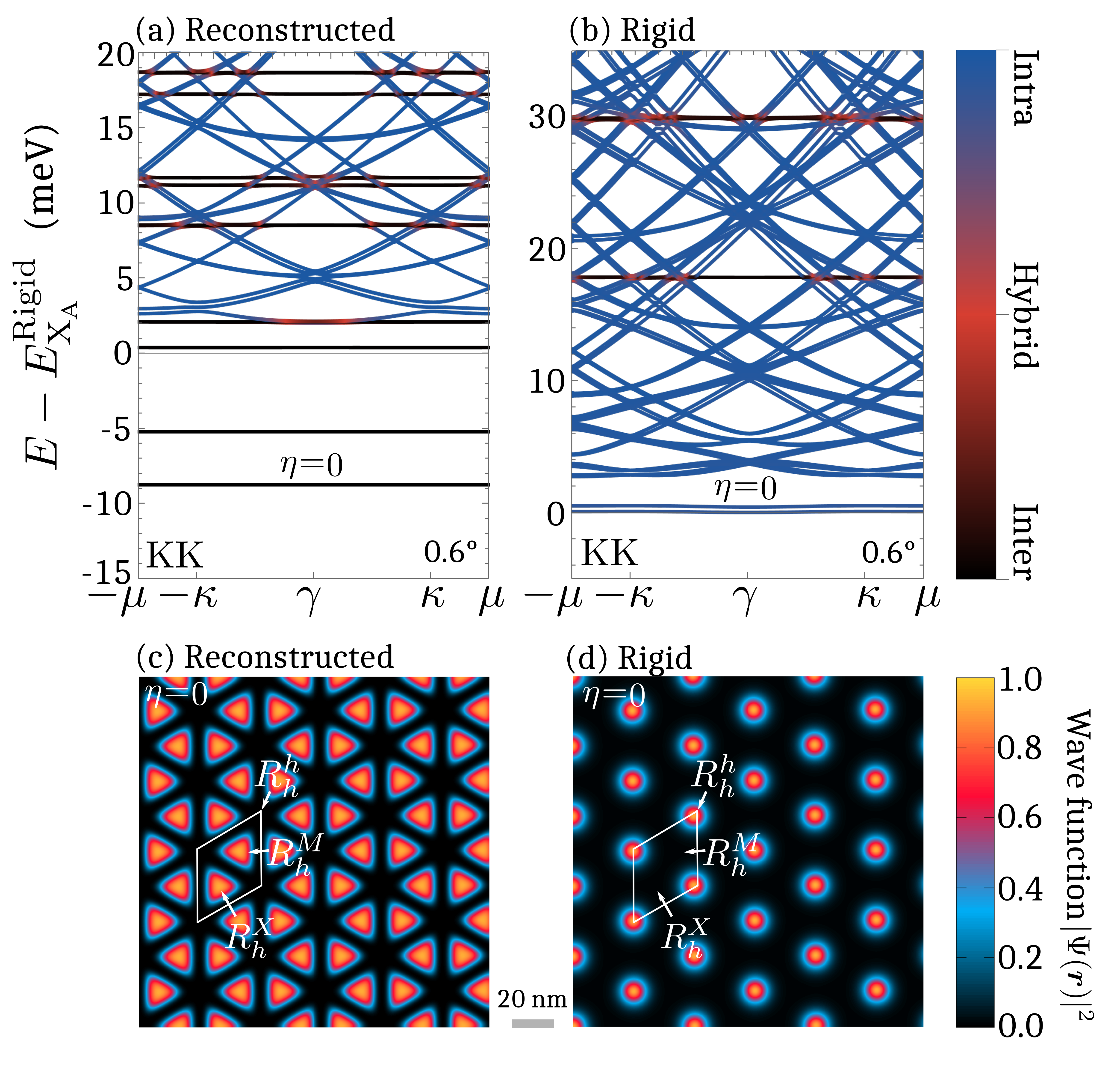}
\caption{\label{fig:5_supp} Band structure and wave function of the bright KK exciton in R-type stacked WSe$_2$ bilayer at $\theta=0.6^{\circ}$ in \textbf{(a,c)} the reconstructed lattice and \textbf{(b,d)} the rigid lattice. Energies are normalized to the lowest-lying KK exciton in the rigid lattice. Note that the reconstruction drastically changes the localization site of excitons and that the efficient moir\'e potential for interlayer excitons shifts them below the intralayer KK.}
\end{figure}

By solving the moir\'e exciton eigenvalue equation in \autoref{eq:eigenvalue} we obtain the moir\'e exciton energy landscape in the presence of atomic reconstruction. In \autoref{fig:5_supp}.a-b we show the moir\'e exciton band structure for R-type stacked WSe$_2$. Similarly to the band structure for H-type shown in the main part of the paper, we find multiple strain-induced flat bands for the interlayer excitons (cf. \autoref{fig:5_supp}.b). However, in contrast to H-type stacking, the intralayer excitons are not subject to any moir\'e potential and will still exhibit a very pronounced curvature. In the rigid lattice, the lowest lying KK state is an intralayer state trapped at the $R^h_h$ sites due to the weak tunneling of carriers at these sites. In the reconstructed lattice, the $R^h_h$ stacking configuration will drastically shrink and thus make the tunneling even less efficient, in turn making the lowest lying KK intralayer state more curved than in the rigid lattice and slightly blue shifted. Furthermore, we also find that the interlayer exciton state becomes the lowest lying in R-type stacking, similarly to H-type stacking, due to the very efficient moi\'e potential for interlayer excitons.

In \autoref{fig:5_supp}.c we show the exciton wave function in the reconstructed lattice for $\theta=0.6^{\circ}$. Here, the wave function is localized at the $R^M_h$ sites with a degenerate exciton at the $R^X_h$ sites due to the symmetry of the potentials (cf. \autoref{fig:4_supp}.a-b). Note that the exciton trapped at $R^M_h$ carries the opposite dipole moment from the exciton trapped at $R^X_h$, implying tunability of localization via an external electrical field \cite{hagel2022electrical}. Additionally, the wave function is following the geometry of the moir\'e potential (cf. \autoref{fig:3_supp}.c). In similarity with H-type stacking, the wave function trapping site is changed in comparison with the rigid lattice. In the rigid lattice the trapping of the wave function is instead determined by the very weak tunneling around the K valley (note that this trapping is very inefficient, which can be observed from the spacing of the bands in \autoref{fig:5_supp}.b) In R-type stacking, the only high symmetry stacking which allows for tunneling around the K valley is $R^h_h$, which consequently traps the excitons in the rigid lattice at this site. This restriction on the tunneling stems from the $C_3$ symmetry of the d-orbitals around the metal atom which mainly composes the orbitals of the K valley \cite{D0NR02160A,PhysRevResearch.3.043217}.  

\subsection{Optical response}
\begin{figure}[t!]
\hspace*{0.0cm}  
\includegraphics[width=1.00\columnwidth]{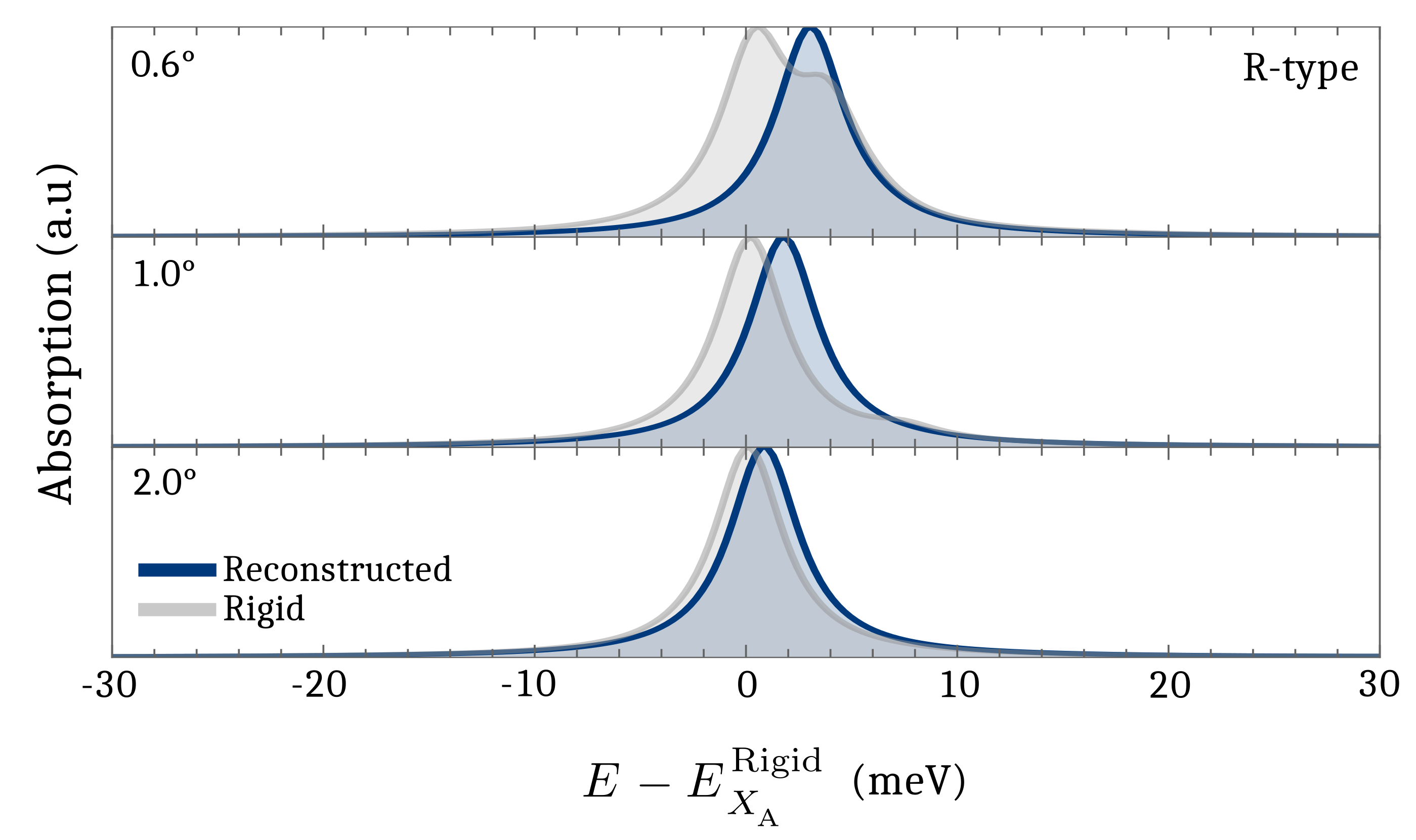}
\caption{\label{fig:6_supp} Absorption spectrum showing the the bright KK exciton for different twist angles in the R-type WSe$_2$ homobilayer. Here, the blue peaks stem from the reconstructed lattice and the gray peaks from the rigid lattice.}
\end{figure}

With the moir\'e exciton band structure for R-type stacked WSe$_2$ (cf. \autoref{fig:5_supp}.a-b) in the reconstructed and rigid lattice regime, we can now calculate the optical response of the material. In \autoref{fig:6_supp} we show the optical absorption for the reconstructed (blue peaks) and the rigid lattice (gray peak). In contrast to H-type stacking where we predicted a very stark difference between the rigid lattice and the reconstructed lattice, in R-type stacking the difference is negligible. This comes from the lack of a strain-induced moir\'e potential in R-type stacking. We observe a slight blue shift of the reconstructed peak at smaller twist angles stemming from the less efficient tunneling of carriers when the $R^h_h$ regions shrink. This also removes the small splitting that otherwise can be found in the rigid lattice (see gray peaks). When increasing the twist angle the peaks align themselves, indicating that the reconstructed lattice is now more rigid. Furthermore, the KK interlayer exciton that does experience a significant moir\'e potential in the reconstructed lattice is not visible in the absorption spectra due to the small oscillator strength.

\subsection{K$\Lambda$ exciton band structure}
\begin{figure}[t!]
\hspace*{0.0cm}  
\includegraphics[width=1.00\columnwidth]{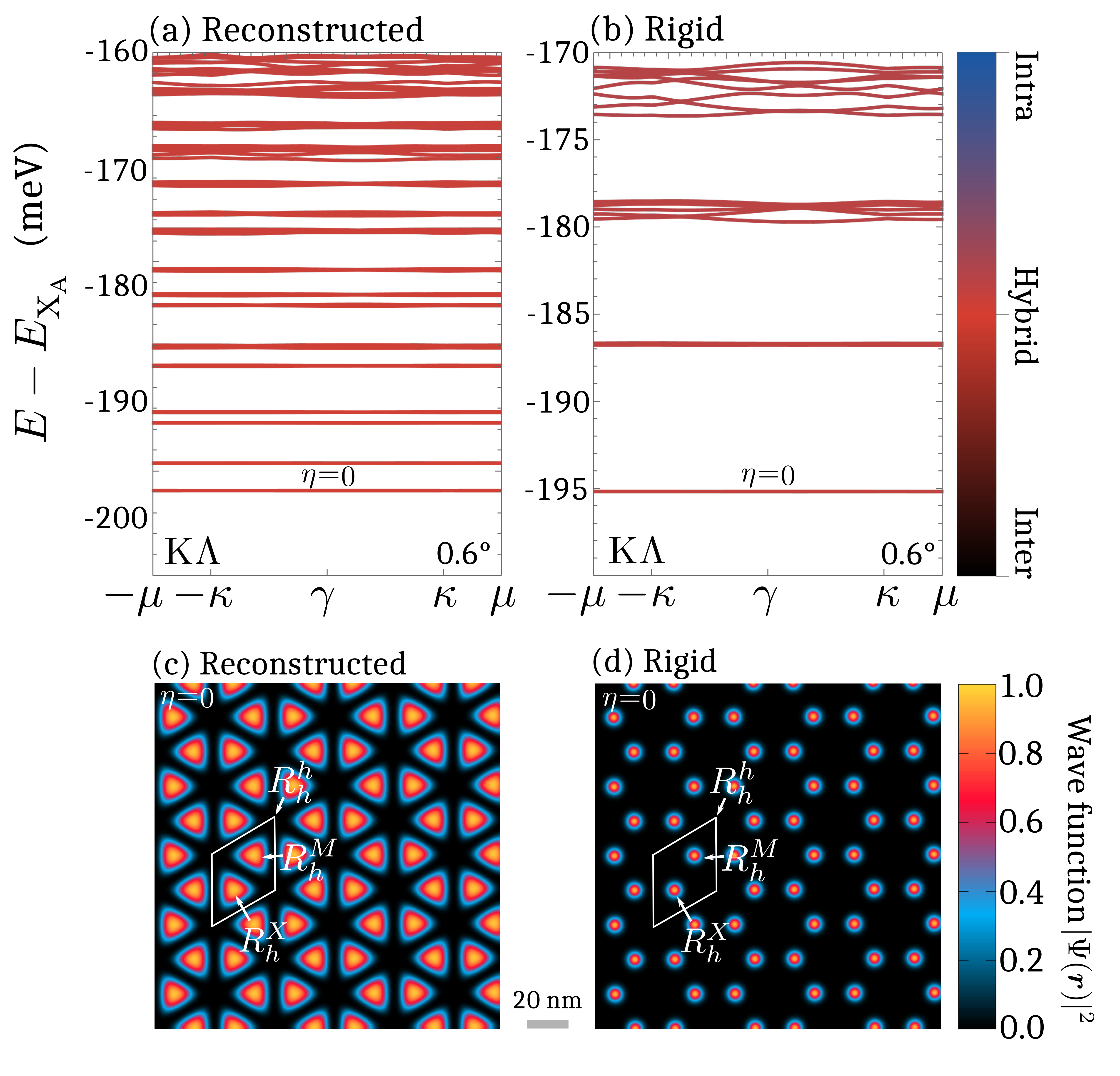}
\caption{\label{fig:7_supp} Band structure of the low-lying dark K$\Lambda$ exciton in R-type stacked WSe$_2$ at $\theta=0.6^{\circ}$ in \textbf{(a)} the reconstructed lattice and \textbf{(b)} the rigid lattice. Energies are normalized to the lowest lying A exciton in their respective lattice. The corresponding K$\Lambda$ exciton wave function for the lowest lying state is shown for \textbf{(c)} the reconstructed and \textbf{(d)} the rigid lattice.}
\end{figure}

In the main part of the paper we discussed the optical response in the form of PL. There, the low-lying K$\Lambda$ exciton is dominating the optical response. Here, we show the corresponding moir\'e exciton band structure for the K$\Lambda$ exciton in R-type stacked WSe$_2$ for both the reconstructed lattice (\autoref{fig:7_supp}.a) and the rigid lattice (\autoref{fig:7_supp}.b). Note that the energies are given with respect to their corresponding A exciton. We notice first the strongly hybridized nature of the K$\Lambda$ exciton in both the rigid lattice and the reconstructed lattice \cite{D0NR02160A}, stemming from the efficient electron tunneling around the $\Lambda$ valley. This consequently means that the dominating component of the moir\'e potential is the tunneling, which is present in both the rigid and reconstructed lattice. However, there are some significant changes between the reconstructed case and the rigid case. Notably, we find far more trapped states in the reconstructed lattice (cf. \autoref{fig:7_supp}.a) in comparison to the rigid lattice (cf. \autoref{fig:7_supp}.b). This is a direct consequence of the additional strain-induced potentials in the reconstructed lattice. In \autoref{fig:7_supp}.c-d we show the K$\lambda$ exciton wave functions for the reconstructed lattice (\autoref{fig:7_supp}.c) and the rigid lattice (\autoref{fig:7_supp}.d). In contrast to the KK exciton wave function, the K$\Lambda$ is already trapped at $R^M_h$($R^X_h$) in the rigid lattice due to the strong tunneling of electrons at these sites. The increased size of these regions, however, increases the size of the wave functions as well, making the K$\Lambda$ exciton more delocalized in the reconstructed lattice than in the rigid lattice.

\section{Theoretical model}
\subsection{Atomic reconstruction}\label{sec:recon}
To obtain the moir\'e exciton landscape in the regime of atomic reconstruction we first need to access the change in relative atomic positions with the twist angle, i.e.  the atom deviation from the rigid lattice site. In a continuum model, this deviation is captured by a displacement vector $\mathbf{u}(\mathbf{r})$, which translates a point in space $\mathbf{r}$ with some distance $\mathbf{u}(\mathbf{r})$. Assuming only small displacements, the displacement vector is then related to the linear strain tensor as $\varepsilon_{ij}=\frac{1}{2}(u_{i,j}+u_{j,i})$, where $u_{i,j}=\frac{1}{2}(\partial_iu_j+\partial_ju_i)$, where $i(j)=(x,y)$. Following the theory of elasticity, the total elastic energy per unit volume can then be written as \cite{landau1986theory}
\begin{equation}
    \mathcal{U}^{l}=\frac{\lambda}{2}(u^{l}_{i,i})^2+\mu u^{l}_{i,j}u^{l}_{j,i},
\end{equation}
where $\lambda$ and $\mu$ are the material specific Lam\'e parameters and $l$ is the layer index. 

We now consider the presence of another layer with a marginal twist angle. Following the approach laid out in Refs.\cite{enaldiev2020stacking,ferreira2021band} we use a parameterized form of the adhesion energy and fit it to data from density functional theory (DFT)
\begin{equation}
\begin{split}
        &W_{R/H}(\mathbf{r}_0)=-\kappa \mathcal{Z}_{R/H}^2(\mathbf{r}_0)\\
        &+\sum_{n=0}^{2}\Big[a_1\text{cos}(\mathbf{G}_n\mathbf{r}_0)+a_2\text{sin}(\mathbf{G}_n\mathbf{r}_0+\gamma_{R/H})\Big],
\end{split}
\end{equation}
where $R/H$ denotes R-type and H-type stacking respectively, and the phase is given by $\gamma_{R}=\pi/2$($\gamma_{H}=0$). Here, $\mathbf{r}_0=\theta \hat{z}\times \mathbf{r}+\mathbf{u}^{t}(\mathbf{r})-\mathbf{u}^{b}(\mathbf{r})$, which means that $\mathbf{G}_n\mathbf{r}_0\approx \mathbf{g}_n\mathbf{r}+\mathbf{G}_n\Delta\mathbf{u}(\mathbf{r})$, where $\Delta \mathbf{u}(\mathbf{r})=\mathbf{u}^{t}(\mathbf{r})-\mathbf{u}^{b}(\mathbf{r})$ and $\mathbf{g}_n$ is the reciprocal vector of the mini-Brillouin zone (mBZ). 

Furthermore, $\mathcal{Z}_{R/H}(\mathbf{r}_0)$ is the deviation of the interlayer distance from $d_0=0.69$ nm \cite{ferreira2021band}, given by
\begin{equation}
\begin{split}
        &Z_{R/H}(\mathbf{r}_0)=\frac{1}{2\kappa}\sum_{n=0}^{2}\Big[a_1A\text{cos}(\mathbf{G}_n\mathbf{r}_0)\\
        &+a_2|\mathbf{G}_n|\text{sin}(\mathbf{G}_n\mathbf{r}_0+\gamma_{R/H})\Big].
\end{split}
\end{equation}
The parameters $\kappa$, $a_1$, $a_2$ and $A$ are all fitted from DFT simulations and are obtained from \cite{ferreira2021band}. The parameters are also given in \autoref{tab:param}. 

\begin{table}[b!]
\centering
\caption{\label{tab:param} Fitted parameters for the adhesion energy and the Lam\'e parameters of bilayer WSe$_2$ obtained from Ref. \cite{ferreira2021band}.}
\begin{tabular}{c|c|c|c|c|c}
\hline
\hline
$a_1$              & $a_2$              & $\kappa$           & A    & $\mu$     & $\lambda$         \\
eV$\cdot$nm$^{-2}$ & eV$\cdot$nm$^{-2}$ & eV$\cdot$nm$^{-4}$ & nm$^{-1}$ & N/m & N/m \\ \hline
0.1340             & 0.0261             & 190                & 29.889  & 48.4 &  29.7    \\ \hline
\hline
\end{tabular}
\end{table}
The total energy can then by described with the following functional integral \cite{ferreira2021band,carr2018relaxation}
\begin{equation}
    \mathcal{E}=\int_{\mathcal{A}_{\text{M}}} d^2\mathbf{r}\Big[\sum_{l}\mathcal{U}^{l}+W_{R/H}(\mathbf{r}_0)\Big],
\end{equation}
where $\mathcal{A}_{\text{M}}$ is the moir\'e unit cell area and $\mathbf{r}$ is the real space coordinate in the moir\'e lattice. To find the relevant displacement vectors $\mathbf{u}^{l}(\mathbf{r})$, we expand them as a Fourier series $\mathbf{u}^{l}(\mathbf{r})=\sum_n \mathbf{u}^{l}_ne^{i\mathbf{g}_n\mathbf{r}}$ and turn the problem into an optimization problem, which can then be solved numerically for the Fourier coefficients \cite{ferreira2021band,carr2018relaxation}. This is done for the WSe$_2$ homobilayer up to the 12th moir\'e shell of the mBZ lattice vectors.    

\subsection{Strain in the reconstructed regime}\label{sec:strain}
The displacement vectors obtained from the optimization problem described in the previous section can be related for small displacements to the strain in the material via the linear strain tensor
\begin{equation}
    \varepsilon^l_{ij}=
    \begin{pmatrix}
       u^l_{x,x} & \frac{1}{2}(u^l_{x,y}+u^l_{y,x})\\
        \frac{1}{2}(u^l_{y,x}+u^l_{x,y}) & u^l_{y,y}
\end{pmatrix}.
\end{equation}
In our considered material, this is a second-rank tensor belonging to the D3h symmetry group. Consequently, in 2D $u^l_{i,j}$ transforms according to scalar component $u^l_{x,x}+u^l_{y,y}$ and a vector component $(u^l_{x,x}-u^l_{y,y},-2u^l_{x,y})$ \cite{amorim2016novel,cazalilla2014quantum,fang2018electronic}. Since we are interested in the effect reconstruction has on the exciton energy landscape, we are mainly interested in what impact the strain has on electronic bands. For this purpose, we can associate the scalar component as uniaxial strain in each direction and the vector component as a vector gauge potential \cite{rostami2015theory}, also known as piezo potential \cite{enaldiev2020stacking,Enaldiev_2021}.

\textbf{Scalar strain potential:}
The band shifts due to the scalar component of the strain are obtained by exploiting the linear dependence of the band on the uniaxial strain \cite{khatibi2018impact}
\begin{equation}\label{eq:scalarStrain}
    S^{l}_{\xi\lambda}(\mathbf{r})=\sum_{i} u^l_{i,i}(\mathbf{r})g^{l}_{\xi \lambda},
\end{equation}
where $\xi$ is the valley index, $\lambda=(c,v)$  the band index, and $g^{l}_{\xi \lambda}$ the valley-specific gauge factor obtained from the DFT calculations preformed in Ref.\cite{khatibi2018impact} and given in \autoref{tab:gauge}. 

\begin{table}[b!]
\centering
\caption{\label{tab:gauge} Linear strain gauge factors for WSe$_2$ for single-particle bands, obtained from DFT calculations Ref.\cite{khatibi2018impact}.}
\begin{tabular}{c|c|c|c|c|c}
\hline
\hline
           & K       & K$^{\prime}$ & $\Lambda$ & $\Lambda^{\prime}$ & $\Gamma$ \\ \hline
Conduction (meV) & -126.13 & -126.13      & -30.33    & -30.33             & -        \\ \hline
Valance (meV)    & -57.50 & -57.50      & -         & -                  & -4.23    \\ \hline
\hline
\end{tabular}
\end{table}

\textbf{Piezo potential:}
The relationship between the polarization $\mathcal{P}^{l}_i$ and the linear strain tensor is given by \cite{bernardini1997spontaneous}
\begin{equation}
    \mathcal{P}^{l}_i=\sum_{jk}u^l_{j,k} e^{l}_{ijk} ,
\end{equation}
where $e^{l}_{ijk}$ is the piezoelectric tensor component. Similarly to the strain tensor, the piezoelectric tensor belongs to D3h and will thus only have certain components contributing. These are \cite{duerloo2012intrinsic}
\begin{equation}
    e^{l}_{11}=e^{l}_{111}=-e^{l}_{122}=-e^{l}_{212}=-e^{l}_{221},
\end{equation}
where we have dropped the third index and only use $e^{l}_{11}$ as the piezo coefficient from now on. The resulting polarization is the vector component of the linear strain tensor established in earlier multiplied by the piezo coefficient
\begin{equation}
    \mathbf{\mathcal{P}}^l(\mathbf{r})=e^l_{11}(u^l_{x,x}-u^l_{y,y},-2u^l_{x,y}).
\end{equation}
By aligning the y-axis of the polarization with the vector from the metal atom to the chalcogen atom we have $e^0_{11}=-e^1_{11}>0$ for H-type configuration and $e^0_{11}=e^1_{11}$ for R-type configuration \cite{enaldiev2020stacking}. The value of the piezo coefficient $e_{11}=2.03\cdot 10^{-10}\text{C/m}$ is obtained from Ref.\cite{rostami2018piezoelectricity}. Note that there is no polarization in uniformly strained lattices without a shear component. Via Gauss law one can write the charge distribution with respect to the polarization
\begin{equation}
    \rho^l_{\text{piezo}}(\mathbf{r})=-\nabla_{\mathbf{r}}\cdot \mathbf{\mathcal{P}}^l(\mathbf{r}).
\end{equation}
By including the out-of-plane direction it reads
\begin{equation}
    \rho^l_{\text{piezo}}=-e_{11}^l\Big[2\partial_x u^l_{x,y}+\partial_y(u^l_{x,x}-u^l_{y,y})\Big]\delta(z-z_l).
\end{equation}
In addition to the piezo charge density, there is also the screening-induced charge density given by the Poisson equation
\begin{equation}
     \frac{\rho_{\text{ind}}}{\alpha^l_{\parallel}}=\delta(z-z^l)\nabla^{2}_{\mathbf{r}}\phi(\mathbf{r},z),
\end{equation}
where $\phi(\mathbf{r},z)$ is the electrostatic potential produced by the piezo charges and $\alpha^l_{\parallel}=d_0\epsilon_0(1-\epsilon^l_{\parallel})$ is the in-plane polarizability \cite{enaldiev2020stacking}. Here, $\epsilon^l_{\parallel}$ is the in-plane dielectric screening obtained from Ref.\cite{laturia2018dielectric}. The total charge density $\rho_{\text{tot}}=\rho_{\text{ind}}+\rho_{\text{piezo}}$ and the piezoelectric potential can then be found by solving the full Poisson equation, which is done by expanding the piezo potential as a Fourier expansion $\phi(\mathbf{r},z)=\sum_{n}\tilde{\phi}_n(z)e^{i\mathbf{g}_n\cdot\mathbf{r}}$ and solving for matching boundary conditions of the two dielectric slabs \cite{enaldiev2020stacking}
\begin{equation}
    [\partial_{zz}^2+\nabla^2_{\mathbf{r}}]\phi(\mathbf{r},z)=(\rho^{l=0}_{\text{tot}}+\rho^{l=1}_{\text{tot}})/\epsilon_0.
\end{equation}
The piezo-induced energy shifts of a charge carrier $P^{\lambda}_{l}(\mathbf{r})$ is then directly obtained from the piezo potential for each layer $\phi_l(\mathbf{r})$. 

\subsection{Derivation of the moir\'e exciton Hamiltonian}
We obtain microscopic access to the moir\'e exciton energy landscape by first considering a Hamiltonian in second quantization, starting in a decoupled monolayer basis \cite{brem2020tunable,D0NR02160A,PhysRevResearch.3.043217,hagel2022electrical}. Here, we take into account the strong Coulomb interaction by solving the generalized Wannier equation \cite{ovesen2019interlayer} and add the moir\'e potential as periodic modifications to these energies. In the regime of atomic reconstruction we will have the two strain induced potentials including  the scalar strain potential $S(\mathbf{r})$ (see \autoref{eq:scalarStrain}) and the vector strain potential, which we will denote as piezo potential $P(\mathbf{r})$. 

In addition to the strain induced potentials, we will also have the components that already exist in the rigid lattice. These are the stacking-dependent alignment shift $V(\mathbf{r})$ \cite{brem2020tunable,PhysRevResearch.3.043217} and the tunneling of charge carriers $T(\mathbf{r})$ \cite{D0NR02160A,PhysRevResearch.3.043217,hagel2022electrical,ruiz2019interlayer,cappelluti2013tight}. The stacking dependent alignment shift $V(\mathbf{r})$ stems from a charge transfer-induced polarization between the layers \cite{tong2020interferences}, which consequently give rise to an electrostatic potential that will be dependent on the relative atomic alignment between the layers. Since this spontaneous charge polarization can only happen if we lack inversion symmetry, it will only be present in R-type stacked configurations \cite{tong2020interferences}. The interlayer tunneling $T(\mathbf{r})$ stems from the wave function overlap between the layers giving rise to a mixing of monolayer eigenstates, which can be interpreted as a tunneling of electrons and holes between the layers \cite{alexeev2019resonantly,gerber2019interlayer,ruiz2019interlayer,D0NR02160A,PhysRevResearch.3.043217}. Since the wave function overlap is largely dependent on the interlayer distance, the tunneling rate is also periodic when introducing a twist angle, due to the variation in interlayer distance throughout the supercell \cite{ruiz2019interlayer,cappelluti2013tight,PhysRevResearch.3.043217,MoireExcitonsLinderalv}. Thus, in the reconstructed regime we have four different components of the moir\'e potential. The corresponding moir\'e Hamiltonian in real space then reads  
\begin{equation}
    \begin{split}
            H_M&=\sum_{\substack{h\lambda\mathbf{r}}}S^{\lambda}_{hh}(\mathbf{r})\Psi^{\lambda\dagger}_h(\mathbf{r})\Psi^{\lambda}_h(\mathbf{r})\\
            &+\sum_{\substack{h\lambda\mathbf{r}}}P^{\lambda}_{hh}(\mathbf{r})\Psi^{\lambda\dagger}_h(\mathbf{r})\Psi^{\lambda}_h(\mathbf{r})\\
            &+\sum_{\substack{h\lambda\mathbf{r}}}V^{\lambda}_{hh}(\mathbf{r})\Psi^{\lambda\dagger}_h(\mathbf{r})\Psi^{\lambda}_h(\mathbf{r})\\
            &+\sum_{\substack{h\neq h^{\prime}\\\lambda\mathbf{r}}}T^{\lambda}_{h h^{\prime}}(\mathbf{r})\Psi^{\lambda\dagger}_h(\mathbf{r})\Psi^{\lambda}_{h^{\prime}}(\mathbf{r})+h.c,
    \end{split}
\end{equation}
where $h^{(\prime)}=(l,\xi)$ is a compound index, taking into account the layer index $l$ and the valley index $\xi$. Furthermore, $\Psi^{(\dagger)}(\mathbf{r})$ are the annihilation (creation) operators in real space. In the effective mass approximation, we can approximate the wave functions close to the high symmetry points in the Brillouin zone (BZ) as a plane wave expansion $\Psi^{\lambda\dagger}_h(\mathbf{r})=\sum_{\mathbf{k}}e^{i\mathbf{k}\cdot\mathbf{r}}\lambda^{\dagger}_{h,\mathbf{k}}$, where $\mathbf{k}$ is the momentum of the electron/hole. In momentum space, the Hamiltonian will thus have the following form
\begin{align}
\begin{split}\label{eq:ElectronHoleHam}
    H_M&=\sum_{\substack{h\lambda\\\mathbf{k}\mathbf{g}}}s^{\lambda}_{hh}(\mathbf{g})\lambda^{\dagger}_{h,\mathbf{k}+\mathbf{g}}\lambda_{h,\mathbf{k}}\\
    &+\sum_{\substack{h\lambda\\\mathbf{k}\mathbf{g}}}p^{\lambda}_{hh}(\mathbf{g})\lambda^{\dagger}_{h,\mathbf{k}+\mathbf{g}}\lambda_{h,\mathbf{k}}\\
    &+\sum_{\substack{h\lambda\\\mathbf{k}\mathbf{g}}}v^{\lambda}_{hh}(\mathbf{g})\lambda^{\dagger}_{h,\mathbf{k}+\mathbf{g}}\lambda_{h,\mathbf{k}}\\
    &+\sum_{\substack{h\neq h^{\prime} \lambda\\ \mathbf{g}\mathbf{k}}}t^{\lambda}_{hh^{\prime}}(\mathbf{g})\lambda^{\dagger}_{h,\mathbf{k}+\mathbf{g}}\lambda_{h^{\prime},\mathbf{k}}+h.c,
    \end{split}
\end{align}
where the matrix elements are given by the Fourier coefficients in the expansion of their real space constituents
\begin{equation}\label{eq:Expansion}
    P^{\lambda}_{hh}(\mathbf{r})=\sum_{\mathbf{g}}p^{\lambda}_{hh}(\mathbf{g})e^{i\mathbf{g}\cdot\mathbf{r}}.
\end{equation}
Here, $p^{\lambda}_{hh}(\mathbf{g})$ is obtained by solving the integral for the Fourier coefficients
\begin{equation}\label{eq:CoefInt}
        p^{\lambda}_{hh}(\mathbf{g})=\frac{1}{\mathcal{A}_{\text{M}}}\int_{\mathcal{A}_{\text{M}}}d\mathbf{r}e^{-i\mathbf{g}\cdot \mathbf{r}}P^{\lambda}_{hh}(\mathbf{r}).
\end{equation}
The area of the moir\'e supercell is denoted as $\mathcal{A}_\text{M}$. Note that we have used the piezo potential here, but the same method to extract the coefficients holds for all four contributions to the moir\'e potential.

In the case of the tunneling and the alignment shift, we can obtain the real space map of the potentials by smoothly interpolating between different high symmetry stackings \cite{PhysRevResearch.3.043217,brem2020tunable,hagel2022electrical}
\begin{equation}\label{eq:FitFormula}
    V^{\lambda}_{hh}(\mathbf{r}_0)=\text{Re}\Big[v^{\lambda}_h+(\mathcal{A}^{\lambda}_h+\mathcal{B}_h^{\lambda}e^{i2\pi/3})\sum_{n=0}^{2}e^{i\mathbf{G}_n\cdot\mathbf{r}_0}\Big],
\end{equation}
where $v^{\lambda}_h$, $\mathcal{A}^{\lambda}_h$ and $\mathcal{B}^{\lambda}_h$ are parameters which are fitted to data from density functional theory Ref.\cite{PhysRevResearch.3.043217}. From \autoref{sec:recon} we have $\mathbf{G}_n\mathbf{r}_0\approx \mathbf{g}_n\mathbf{r}+\mathbf{G}_n\Delta\mathbf{u}(\mathbf{r})$, where $\Delta\mathbf{u}(\mathbf{r})=0$ for the smooth interpolation of the rigid lattice. We then take into account the deformation of the lattice, by allowing the displacement vectors to deform the geometry, in turn yielding
\begin{equation}\label{eq:FitFormula}
\begin{split}
    V^{\lambda}_{hh}(\mathbf{r})=\text{Re}\Big[v^{\lambda}_h+(\mathcal{A}^{\lambda}_h\\
    +\mathcal{B}_h^{\lambda}e^{i2\pi/3})\sum_{n=0}^{2}e^{i(\mathbf{g}_n\cdot\mathbf{r}+\mathbf{G}_n\cdot \Delta\mathbf{u}^l(\mathbf{r}))}\Big].
\end{split}
\end{equation}

After obtaining the full moir\'e Hamiltonian in electron-hole basis, we can now transform it into exciton basis by considering the pair operator expansion. Under the assumption of low densities, we can expand the single particle operators in pairs $D^{\dagger}_{hk,h^{\prime}\mathbf{k}^{\prime}}=c^{ \dagger}_{h \mathbf{k}}v_{h^{\prime} \mathbf{k}^{\prime}}$. Consequently, we obtain the following expression \cite{katsch2018theory,D0NR02160A,brem2020tunable,hagel2022electrical}
\begin{equation}\label{eq:pairoperator}
\begin{split}
c^{\dagger}_{h \mathbf{k}}c_{h^{\prime} \mathbf{k}^{\prime}}\approx\sum_{m\mathbf{p}}D^{\dagger}_{h\mathbf{k},m\mathbf{p}}D_{h^{\prime}\mathbf{k}^{\prime},m\mathbf{p}}\\
v^{\dagger}_{h \mathbf{k}}v_{h^{\prime} \mathbf{k}^{\prime}}\approx\delta^{hh^{\prime}}_{\mathbf{k}\mathbf{k}^{\prime}}-\sum_{m\mathbf{p}}D^{\dagger}_{m\mathbf{p},h^{\prime}\mathbf{k}^{\prime}}D_{m\mathbf{p},h\mathbf{k}}.
 \end{split}
\end{equation}
We can now transform into exciton basis by expanding the pair operators in a set of orthogonal basis functions
\begin{equation}
D^{\dagger}_{h\mathbf{k},h^{\prime}\mathbf{k}^{\prime}}=\sum_{\mu}X^{\mu\dagger}_{hh^{\prime},\mathbf{k}-\mathbf{k}^{\prime}+\xi_h-\xi_h^{\prime}}\Psi^{\mu}_{hh^{\prime}}(\alpha_{hh^{\prime}}\mathbf{k}^{\prime}+\beta_{hh^{\prime}}\mathbf{k}),
\end{equation}
where $\alpha_{hh^{\prime}}(\beta_{hh^{\prime}})=m^{c(v)}_{h(h^{\prime})}/(m^c_h+m^v_{h^{\prime}})$, $\xi_h$ is the valley index of the compound $h$ and $\mu$ is the exciton quantum number, which in this work is restricted to be 1s. Furthermore, $\Psi^{\mu}_{hh^{\prime}}(\mathbf{k})$ is the exciton wave function that solves the Wannier equation \cite{ovesen2019interlayer}. Moreover, $X^{(\dagger)}$ are the exciton annihilation (creation) operators. Performing the above transformation now allows us to write the complete moir\'e exciton Hamiltonian  
\begin{equation}\label{eq:ExcitonHam}
\begin{split} H_0&=\sum_{L\mathbf{Q}\mathbf{\xi}}E^{\mathbf{\xi}}_{L\mathbf{Q}}X^{\mathbf{\xi}\dagger}_{L,\mathbf{Q}}X^{\mathbf{\xi}}_{L,\mathbf{Q}}\\
    &+\sum_{L\mathbf{Q}\mathbf{\xi}\mathbf{g}}M^{\mathbf{\xi}}_{L\mathbf{g}}X^{\mathbf{\xi}\dagger}_{L,\mathbf{Q}+\mathbf{g}}X^{\mathbf{\xi}}_{L,\mathbf{Q}}\\
&+\sum_{LL^{\prime}\mathbf{Q}\mathbf{\xi}\mathbf{g}}T^{\mathbf{\xi}}_{LL^{\prime}\mathbf{g}}X^{\mathbf{\xi}\dagger}_{L,\mathbf{Q}+\mathbf{g}}X^{\mathbf{\xi}}_{L^{\prime},\mathbf{Q}} +h.c.
    \end{split}
\end{equation}
Here, $L=(l_e,l_h)$ is a compound layer index, $\mathbf{Q}$ is the center-of-mass momentum and $\mathbf{\xi}=(\xi_e,\xi_h)$ is the compound exciton valley index. Additionally, we have introduced the notation $M^{\mathbf{\xi}}_{L\mathbf{g}}$ for the total layer-diagonal moir\'e potential given by $M^{\mathbf{\xi}}_{L\mathbf{g}}=S^{\mathbf{\xi}}_{L\mathbf{g}}+P^{\mathbf{\xi}}_{L\mathbf{g}}+V^{\mathbf{\xi}}_{L\mathbf{g}}$, which includes the scalar strain potential $S^{\mathbf{\xi}}_L(\mathbf{g})$, the piezo potential $P^{\mathbf{\xi}}_L(\mathbf{g})$ and the alignment shift $V^{\mathbf{\xi}}_L(\mathbf{g})$. Furthermore, $E^{\mathbf{\xi}}_{L\mathbf{Q}}$ is the exciton dispersion given by
\begin{equation}\label{eq:dispersion}
    E^{\mathbf{\xi}}_{L\mathbf{Q}}=\hbar^2\frac{(\mathbf{Q}-[\xi_e-\xi_h])^2}{2[m_e+m_h]}+\varepsilon^c_{\xi_e0}-\varepsilon^v_{\xi_h0}+E^\text{b}_{\mathbf{\xi}},
\end{equation}
where $E^\text{b}_{\mathbf{\xi}}$ are the exciton binding energies as obtained from the generalized Wannier equation \cite{ovesen2019interlayer} and $\varepsilon^{\lambda}_{\mathbf{\xi}_{\lambda}}$ is the valley splitting. Effective electron (hole) masses and valley splittings are obtained from Ref. \cite{Korm_nyos_2015}. The matrix elements of the Hamiltonian \autoref{eq:ExcitonHam} then read
\begin{equation}
\begin{split}
&M^{\mathbf{\xi}}_L(\mathbf{g})=m^c_{l_{e}}(\mathbf{g})\mathcal{F}^{\mathbf{\xi}}_{LL}(\beta_{LL}\mathbf{g})-m^v_{l_{h}}(\mathbf{g})\mathcal{F}^{\mathbf{\xi}*}_{LL}(-\alpha_{LL}\mathbf{g}),\\
&T^{\mathbf{\xi}}_{LL^{\prime}}(\mathbf{g})=\Big[\delta_{l_h,l_h^{\prime}}(1-\delta_{l_e,l_e^{\prime}})t^{c\mathbf{\xi}_{e}}_{l_el_e^{\prime}}(\mathbf{g})\mathcal{F}^{\mathbf{\xi}}_{LL^{\prime}}(\beta_{LL^{\prime}}\mathbf{g})\\
&-\delta_{l_e,l_e^{\prime}}(1-\delta_{l_h,l_h^{\prime}})t^{v\mathbf{\xi}_h}_{l_hl_h^{\prime}}(\mathbf{g})\mathcal{F}^{*\mathbf{\xi}}_{LL^{\prime}}(-\alpha_{LL^{\prime}}\mathbf{g})\Big],
\end{split}
\end{equation}
where $m^{\lambda}_{l_{\lambda}}(\mathbf{g})=s^{\lambda}_{l_{\lambda}}+p^{\lambda}_{l_{\lambda}}+v^{\lambda}_{l_{\lambda}}$ is the short notation for the layer-diagonal Fourier coefficients, while $t^{\lambda\mathbf{\xi}_{\lambda}}_{l_{\lambda}l_{\lambda}^{\prime}}$ describes the tunneling Fourier coefficients, both obtained from \autoref{eq:CoefInt}. Here, $\mathcal{F}^{\mathbf{\xi}}_{LL^{\prime}}(\mathbf{q})$ are the form factors given by $\mathcal{F}^{\mathbf{\xi}}_{LL^{\prime}}(\mathbf{q})=\sum_{\mathbf{k}}\Psi^{\mathbf{\xi}*}_{L}(\mathbf{k})\Psi^{\mathbf{\xi}}_{L^{\prime}}(\mathbf{k}+\mathbf{q})$.

\subsection{Diagonalization of moir\'e exciton Hamiltonian}
Once the complete moir\'e exciton Hamiltonian in the reconstructed regime (see \autoref{eq:ExcitonHam}) is developed, we now find a diagonal form of this Hamiltonian. We do this by applying the well-known zone-folding scheme \cite{brem2020tunable,D0NR02160A,hagel2022electrical}. Here, we restrict the summation over the center-of-mass momentum $\mathbf{Q}$ to the first mBZ and then fold it back in with the mBZ lattice vectors $\mathbf{g}$
\begin{equation}
\begin{split}
    H_0&=\sum_{\substack{L\mathbf{Q}\mathbf{\xi}\\\mathbf{g}}}E^{\mathbf{\xi}}_{L\mathbf{Q}}(\mathbf{g})X^{\mathbf{\xi}\dagger}_{L,\mathbf{Q}+\mathbf{g}}X^{\mathbf{\xi}}_{L,\mathbf{Q}+\mathbf{g}}\\
    &+\sum_{\substack{L\mathbf{Q}\mathbf{\xi}\\\mathbf{g}\mathbf{g}^{\prime}}}M^{\mathbf{\xi}}_L(\mathbf{g}^{\prime})X^{\mathbf{\xi}\dagger}_{L,\mathbf{Q}+\mathbf{g}+\mathbf{g}^{\prime}}X^{\mathbf{\xi}}_{L,\mathbf{Q}+\mathbf{g}}\\
    &+\sum_{\substack{LL^{\prime}\mathbf{Q}\\\mathbf{\xi}\mathbf{g}\mathbf{g}^{\prime}}}T^{\mathbf{\xi}}_{LL^{\prime}}(\mathbf{g}^{\prime})X^{\mathbf{\xi}\dagger}_{L^{\prime},\mathbf{Q}+\mathbf{g}+\mathbf{g}^{\prime}}X^{\mathbf{\xi}}_{L,\mathbf{Q}+\mathbf{g}}+h.c,
       \end{split}
\end{equation}
where $\mathbf{Q}\in$mBZ and the notation $E^{\mathbf{\xi}}_{L\mathbf{Q}}(\mathbf{g})=E^{\mathbf{\xi}}_{L\mathbf{Q}+\mathbf{g}}$. By changing to the zone-folding operator basis $F_{L\mathbf{Q}\mathbf{g}}^{\mathbf{\xi}}=X^{\mathbf{\xi}}_{L,\mathbf{Q}+\mathbf{g}}$ we get the following expression
\begin{equation}\label{eq:zonefoldedHam}
    \begin{split}
        H_0&=\sum_{\substack{L\mathbf{Q}\mathbf{\xi}\\\mathbf{g}}}E^{\mathbf{\xi}}_{L\mathbf{Q}}(\mathbf{g})F_{L\mathbf{Q}\mathbf{g}}^{\mathbf{\xi}\dagger}F_{L\mathbf{Q}\mathbf{g}}^{\mathbf{\xi}}\\
        &+\sum_{\substack{L\mathbf{Q}\mathbf{\xi}\\\mathbf{g}\mathbf{g}^{\prime}}}M^{\mathbf{\xi}}_L(\mathbf{g},\mathbf{g}^{\prime})F_{L\mathbf{Q}\mathbf{g}^{\prime}}^{\mathbf{\xi}\dagger}F_{L\mathbf{Q}\mathbf{g}}^{\mathbf{\xi}}\\
        &+\sum_{\substack{LL^{\prime}\mathbf{Q}\\\mathbf{\xi}\mathbf{g}\mathbf{g}^{\prime}}}T^{\mathbf{\xi}}_{LL^{\prime}}(\mathbf{g},\mathbf{g}^{\prime})F_{L\mathbf{Q}\mathbf{g}^{\prime}}^{\mathbf{\xi}\dagger}F_{L\mathbf{Q}\mathbf{g}}^{\mathbf{\xi}}+h.c.
    \end{split}
\end{equation}
Here, we have introduced the abbreviation $M^{\mathbf{\xi}}_L(\mathbf{g},\mathbf{g}^{\prime})=M^{\mathbf{\xi}}_L(\mathbf{g}^{\prime}-\mathbf{g})$.

We now introduce the hybrid moir\'e exciton basis, where we expand the zone-folded operators with a set of orthogonal basis functions called mixing coefficients \cite{brem2020tunable,D0NR02160A,PhysRevResearch.3.043217,hagel2022electrical}
\begin{equation}
    Y^{\dagger}_{\mathbf{\xi}\eta\mathbf{Q}}=\sum_{\mathbf{g}L}\mathcal{C}^{\mathbf{\xi}\eta*}_{L\mathbf{g}}(\mathbf{Q})F^{\mathbf{\xi}\dagger}_{L\mathbf{Q}\mathbf{g}},
\end{equation}
where $\eta$ is the new moir\'e exciton band index and $\mathcal{C}^{\mathbf{\xi}\eta}_{L\mathbf{g}}(\mathbf{Q})$ are the mixing coefficients that fulfill the following conditions
\begin{equation}
\begin{split}
    \sum_{L\mathbf{g}}\mathcal{C}^{\mathbf{\xi}\eta_1*}_{L\mathbf{g}}(\mathbf{Q})\mathcal{C}^{\mathbf{\xi}\eta_2}_{L\mathbf{g}}(\mathbf{Q})=\delta_{\eta_1\eta_2}\\
    \sum_{\eta}\mathcal{C}^{\mathbf{\xi}\eta*}_{L\mathbf{g}}(\mathbf{Q})\mathcal{C}^{\mathbf{\xi}\eta}_{L^{\prime}\mathbf{g}^{\prime}}(\mathbf{Q})=\delta_{LL^{\prime}}\delta_{\mathbf{g}\mathbf{g}^{\prime}}.
\end{split}    
\end{equation}
Applying the above transformation to \autoref{eq:zonefoldedHam} we reveal the moir\'e exciton eigenvalue equation
\begin{equation}\label{eq:eigenvalue}
    \begin{split}
        E^{\mathbf{\xi}}_{L\mathbf{Q}}(\mathbf{g})\mathcal{C}^{\mathbf{\xi}\eta}_{L\mathbf{g}}(\mathbf{Q})+\sum_{\mathbf{g}^{\prime}}M^{\mathbf{\xi}}_L(\mathbf{g},\mathbf{g}^{\prime})\mathcal{C}^{\mathbf{\xi}\eta}_{L\mathbf{g}^{\prime}}(\mathbf{Q})\\
        +\sum_{L^{\prime}\mathbf{g}^{\prime}}T^{\mathbf{\xi}}_{LL^{\prime}}(\mathbf{g},\mathbf{g}^{\prime})\mathcal{C}^{\mathbf{\xi}\eta}_{L^{\prime}\mathbf{g}^{\prime}}(\mathbf{Q})=\mathcal{E}^{\mathbf{\xi}}_{\eta\mathbf{Q}}\mathcal{C}^{\mathbf{\xi}\eta}_{L\mathbf{g}}(\mathbf{Q}),
    \end{split}   
\end{equation}
where $\mathcal{E}^{\mathbf{\xi}}_{\eta\mathbf{Q}}$ are the final moir\'e exciton energies, which can be obtained by numerically solving the equation above. Consequently, we obtain the diagonal form of the moir\'e exciton Hamiltonian
\begin{equation}
    H_0=\sum_{\mathbf{Q}\mathbf{\xi}\eta}\mathcal{E}^{\mathbf{\xi}}_{\eta\mathbf{Q}}Y^{\dagger}_{\mathbf{\xi}\eta\mathbf{Q}}Y_{\mathbf{\xi}\eta\mathbf{Q}}.
\end{equation}

\subsection{Optical response}
Our approach for the optical response of the material is done in the exact way as described in the supplementary of Ref. \cite{hagel2022electrical}. Here, we use the photoluminescence (PL) formula for phonon-assisted PL first derived in Refs.\cite{PLBrem,D0NR02160A}, thus allowing us to obtain the optical response of both bright and momentum dark excitons.